\documentclass[conference]{IEEEtran}
\IEEEoverridecommandlockouts

\usepackage{cite}
\usepackage{amsmath,amssymb,amsfonts}
\usepackage{algpseudocode}
\usepackage{algorithm}
\usepackage{float}
\usepackage{graphicx}
\usepackage{textcomp}
\usepackage{xcolor}
\def\BibTeX{{\rm B\kern-.05em{\sc i\kern-.025em b}\kern-.08em
    T\kern-.1667em\lower.7ex\hbox{E}\kern-.125emX}}
\begin{document}

\title{Interference-Free RIS-Aided Cell-Free Massive MIMO with Physical Layer Security\\
}

\author{
    \IEEEauthorblockN{
        Sumeyra Hassan,~\IEEEmembership{Graduate Student Member,~IEEE}, Bin Li, Yalcin Sadi,~\IEEEmembership{Member,~IEEE}\\ 
        Erdal Panayırcı, ~\IEEEmembership{Life Fellow,~IEEE}, H. Vincent Poor,~\IEEEmembership{Life Fellow,~IEEE}
    }
    \thanks{This research has been supported by the Bilateral Scientific Cooperation Program between the U.S. National Science Foundation (NSF) and the Scientific and Technical Research Council of Turkiye (TÜBİTAK), Project No. 123N122, Turkiye.}
\thanks{S. Hassan, Y.  Şadi, and E. Panayirci are with the Department of Electrical and Electronics Engineering, Kadir Has University, 34083 Istanbul, Turkiye (e-mail: sumeyra.aldemir@khas.edu.tr; yalcin.sadi@khas.edu.tr; eepanay@khas.edu.tr). }
 \thanks{B. Li is with the School of Electronic Information, Wuhan University, Wuhan 430072, China, and also with the 6G Intelligent Connectivity International Science and Technology Cooperation Center, Wuhan 441800, China (e-mail: binli6@whu.edu.cn).}
\thanks{H. Vincent Poor is with the Department of Electrical and Computer Engineering, Princeton University, Princeton, NJ 08544 USA (e-mail: poor@princeton.edu).}}

\maketitle

\begin{abstract}
In this paper, a reconfigurable intelligent surface (RIS)–assisted cell-free massive MIMO (CFmMIMO) framework is designed to enhance physical layer security (PLS) and mitigate multi-user (MU) interference in next-generation wireless networks. A channel state information (CSI)–based precoder is designed at the access point (AP) to suppress MU interference, enabling interference-free reception for the legitimate users. To further enhance secrecy performance, we formulate a joint optimization problem that maximizes the secrecy sum rate using an alternating optimization (AO) framework, which iteratively updates the active beamforming at the AP, user power allocation, and the RIS phase-shift matrix. The highly non-convex problem is addressed under the Riemannian manifold optimization (RMO) framework and solved using a Riemannian Conjugate Gradient (RCG) algorithm for RIS phase-shift design. Simulation results verify that the proposed framework effectively enhances the secrecy sum rate and eliminates interference, demonstrating its potential for secure and scalable CFmMIMO networks in dense wireless environments.
\end{abstract}


\section{Introduction}
The rapid growth of wireless communication systems has led to an ever-increasing demand for higher data rates, enhanced security, and ubiquitous coverage. As we transition towards the sixth generation (6G) of wireless networks, meeting these demands becomes increasingly challenging \cite{bir}. Traditional cellular architectures, characterized by centralized base stations (BS) and constrained by limited spectral efficiency, are often inadequate to fulfill the requirements of next-generation wireless communication. This limitation has led to the exploration of innovative solutions, such as cell-free massive MIMO (CFmMIMO) networks, to overcome the inadequacies of current wireless systems \cite{firstCF}. 

CFmMIMO systems comprise a large number of geographically distributed access points (APs) that are implemented and coordinated by a central processing unit (CPU), enabling seamless coverage and high spectral efficiency through coordinated transmission and reception \cite{uniform}. Studies have shown that CFmMIMO can significantly mitigate energy consumption and enhance capacity, offering uniformly high service quality throughout the coverage area \cite{mimovssmall, dort}. Recent advances further demonstrate that secure and energy-efficient CFmMIMO architectures can be achieved by jointly optimizing active and passive components \cite{binliIOS2025,binliFullDuplex2024}, highlighting the growing importance of system-level co-design in dense deployments. In parallel, multi-RIS and massive-MIMO–aided CF architectures have also been shown to significantly enhance connectivity and resource allocation efficiency \cite{binliMultiRIS2024}. Despite these advances, CFmMIMO networks continue to face critical challenges related to interference management and physical layer security (PLS).

Interference is a critical problem in dense wireless networks, as it can significantly degrade the quality of service and overall system performance. For example, \cite{hassanInterference} has demonstrated that precoder-assisted interference mitigation frameworks can effectively suppress multi-user interference by optimizing transmit weights across distributed antennas, thereby enhancing link reliability and improving overall spectral efficiency. In parallel, intelligent reflecting surfaces (IRSs) or reconfigurable intelligent surfaces (RISs) have been introduced as a transformative technology to enhance coverage, spectral efficiency, and link robustness by reconfiguring the wireless propagation environment. By intelligently adjusting the phase shifts of incident signals, RISs can shape the radio channel to strengthen desired signal paths while naturally reducing unintended interference \cite{haraald,irs}. Effective interference suppression remains crucial for fully exploiting the benefits of CFmMIMO. For instance, a comparative study between CFmMIMO and RIS architectures demonstrates that CFmMIMO significantly improves cell-edge performance by coherently suppressing interference, particularly in dense deployment scenarios \cite{yedi}. Furthermore, the use of RIS in CFmMIMO systems with maximum ratio combining (MRC) over spatially correlated channels has been investigated to optimize system performance by effectively configuring RIS elements \cite{sekiz}. The work in \cite{sekiz} demonstrates that an appropriate RIS configuration can significantly improve system throughput and user fairness by leveraging spatial diversity in large-scale deployments.

PLS has gained increasing attention in modern wireless communication systems, particularly with the evolution toward 6G \cite{erdal}. Although traditional cryptographic techniques at the application layer remain vital, they often prove insufficient in dynamic and heterogeneous wireless environments, where adversaries can exploit channel conditions. As a result, PLS has emerged as a complementary strategy that exploits the randomness, fading, and noise characteristics of the wireless medium to ensure confidentiality at the signal level. Various signal processing-based techniques, such as cooperative relaying \cite{on}, artificial noise (AN)-aided beamforming \cite{onbir}, and cooperative jamming \cite{oniki}, have been introduced to enhance secrecy performance; however, these approaches often increase hardware complexity and power consumption. Recent studies have demonstrated that combining transmit precoding and channel state information (CSI)–based signal shaping can substantially strengthen the legitimate user’s link while degrading the signal quality at potential eavesdroppers, thereby achieving secure transmission without relying on additional jamming nodes \cite{hassanSSK,hassanGSSK}. 

To address the challenges of interference management and PLS in densely populated wireless environments, RISs have emerged as a promising, low-cost solution. By intelligently adjusting the phase shifts of the reflected electromagnetic waves, RISs can reshape the propagation environment to enhance legitimate communication links while suppressing undesired interference and signal leakage \cite{irspls}. When integrated into CFmMIMO architectures, RISs provide additional spatial degrees of freedom that can be jointly optimized with active beamforming at distributed access points to improve both spectral efficiency and secrecy performance \cite{irsplscfmimo,secrecyPerformance,nadeem}. Recent architectural advances, such as beyond-diagonal RIS (BD-RIS) designs, further expand these capabilities by enabling inter-element coupling, which facilitates joint control of reflection amplitude and phase for more flexible channel manipulation \cite{xiongBDRIS}. Moreover, comprehensive surveys highlight that RIS-assisted wireless systems inherently unify interference mitigation and secrecy enhancement as key design goals for next-generation secure and energy-efficient 6G networks \cite{khoshafaSurvey}.

Recent advances in optimization theory have demonstrated that Riemannian manifold optimization (RMO) provides a practical framework for addressing the highly non-convex problems that arise in RIS-aided wireless systems \cite{Li2025}. Conventional methods, such as semidefinite relaxation, alternating optimization, or majorization–minimization, typically transform the original problem into relaxed forms, leading to suboptimal solutions and a high computational burden. In contrast, RMO exploits the intrinsic geometric structure of the feasible set, such as the unit-modulus and orthogonality constraints of RIS phase shifts and precoding matrices, to conduct optimization directly on smooth manifolds. This geometric interpretation reformulates constant-modulus constraints as complex circle or sphere manifolds, enabling gradient-based updates that inherently satisfy physical constraints. By unifying active and passive beamforming within a manifold domain, optimization can be performed efficiently without violating feasibility conditions. The unified manifold optimization (UMO) framework introduced in \cite{unifiedManifold} achieves faster convergence and higher spectral efficiency than Euclidean-based techniques by performing parallel conjugate-gradient updates over coupled manifolds. In addition, manifold-based formulations have been proposed for interference mitigation and large-scale MIMO beamforming, confirming their computational stability and efficiency across complex-valued domains \cite{circleManifold,manifoldRISmMIMO,sumRateRCC}. Furthermore, \cite{sumSecrecyISAC} demonstrated that the same geometric principles can be applied to joint active–passive optimization with augmented Lagrangian and alternating-manifold updates, proving that RMO achieves significant performance gains with lower complexity compared to classical convex methods. Collectively, these studies establish RMO as a powerful and general optimization approach for next-generation RIS-aided communication systems.

The main contributions of this paper can be summarized as follows:

\begin{itemize}
\item We develop a CSI-based precoding design that effectively eliminates multi-user (MU) interference in a RIS-assisted CFmMIMO network, enabling interference-free transmission and reliable signal recovery for all legitimate users.

\item By integrating the proposed precoder with RIS-aided beamforming, we enhance physical layer security (PLS) by steering constructive signal energy toward legitimate users while suppressing information leakage to potential eavesdroppers.

\item We formulate a joint secrecy sum rate maximization problem that is solved using an alternating optimization (AO) framework, which iteratively updates the beamforming vectors at the access points, the user power allocation, and the RIS phase shifts.

\item To efficiently solve the RIS phase-shift subproblem, we adopt an RMO framework and implement a Riemannian Conjugate Gradient (RCG) algorithm that exploits the geometric structure of the unit-modulus manifold, ensuring constraint satisfaction and computational efficiency.

\item The analytical and simulation results demonstrate that the proposed framework achieves an effective trade-off among secrecy performance, interference suppression, and computational efficiency, making it suitable for practical RIS-assisted CFmMIMO networks.
\end{itemize}
\section{System Model}
\begin{figure}[b]
\centerline{\includegraphics[width=80mm]{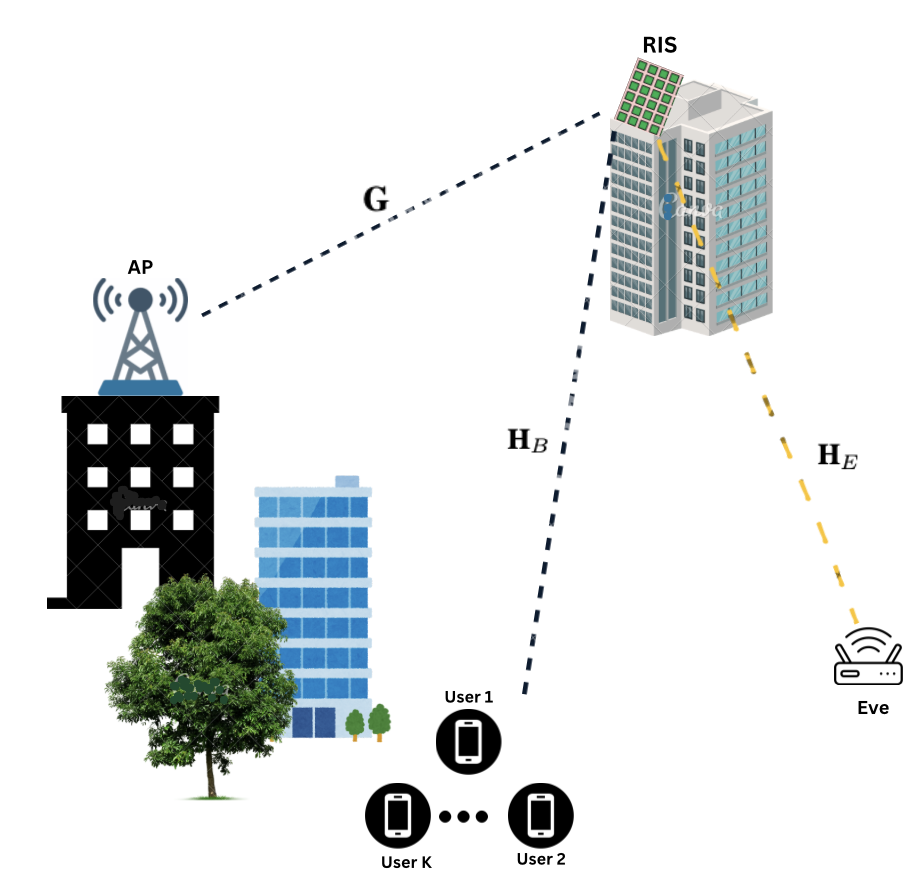}}
\caption{RIS-assisted CFmMIMO communication.}
\label{fig}
\end{figure}
In this paper, we consider the downlink of a RIS and precoding-aided CFmMIMO network, as illustrated in Fig. 1. The network consists of a single AP serving $K$ users (UEs), a single full-duplex (FD) active Eve, and a single RIS. A CPU manages control and planning, with the AP and RIS connected via a wireless backhaul. The AP is equipped with $N_t$ transmit antennas, while each user has a single antenna. The RIS is composed of $M$ reflection elements. Unlike UEs, Eve is equipped with multiple receive antennas ($N_{E} > 1$) and may attempt to intercept the signals intended for legitimate users.

To further optimize network performance, the $N_t$ antennas at the AP are divided into $K$ blocks, with each block containing $n_t = N_t/K$ antennas, where $n_t$ is selected as a power of 2. For the time being, without loss of generality, each block $b_u$ is dedicated to transmitting data to a specific user, allowing the system to serve multiple users simultaneously. Perfect effective CSI of the AP–RIS–UE cascaded channels is assumed to be available at the AP, enabling optimal precoding and beamforming for each user’s data transmission.

The RIS is strategically deployed within the XOY plane at a high altitude to optimize its reflection behavior and mitigate the risk of interception by potential eavesdroppers. It is equipped with $M$ programmable phase shifters, which allow it to enhance signal reflection toward legitimate users while minimizing signal leakage toward the Eve. Given that the direct links between the AP and legitimate users are assumed to be obstructed by physical obstacles (e.g., buildings, trees), the AP communicates with the users through the RIS, directing its transmit beams towards the RIS. The RIS then reflects these beams downward to serve the legitimate users, minimizing the risk of signal interception by Eve. The strategic positioning and beamforming ensure that the reflected signals are effectively directed toward the intended users, reducing the likelihood of interception.

The downlink received signals $\mathbf{y}_B \in \mathbb{C}^{K \times 1}$ from all legitimate users can generally be expressed as
\begin{equation}
    \mathbf{y}_B = \mathbf{H}_B^{\dagger} \boldsymbol{\Theta} \mathbf{G}\mathbf{x} + \mathbf{n}_B,
\end{equation}
where $\mathbf{H}_B \in \mathbb{C}^{M \times K}$ denotes the channel matrix between the RIS and all legitimate users. $\boldsymbol{\Theta} = \mathrm{diag}(\mathbf{v}) \in \mathbb{C}^{M \times M}$ is the RIS phase–shift matrix with $\mathbf{v} = [e^{j\theta_1},e^{j\theta_2},\ldots,e^{j\theta_M}]^T$. The AP–RIS channel is represented by $\mathbf{G} = \big[\mathbf{G}_1 \ \mathbf{G}_2 \ \cdots \ \mathbf{G}_K\big] \in \mathbb{C}^{M \times N_t}$. The transmitted signal is given by 
$\mathbf{x} = \mathbf{W}_{\boldsymbol{\mathcal{P}}} 
S\boldsymbol\varpi \in \mathbb{C}^{N_t \times 1}$, where 
$\mathbf{W}_{\boldsymbol{\mathcal{P}}} =\text{diag}\left(\sqrt{\mathcal{P}_1}\mathbf{w}_1,\sqrt{\mathcal{P}_2}\mathbf{w}_2, \cdots, \sqrt{\mathcal{P}_K}\mathbf{w}_K\right) \in \mathbb{C}^{N_t \times K}$ 
is a beamforming matrix containing the beamforming vectors $\mathbf{w}_k \in \mathbb{C}^{n_t \times 1}$ for each user $k$, with $\|\mathbf{w}^{\dagger}_k \mathbf{w}_k\|^2=1$, $\boldsymbol{\mathcal{P}}=[\mathcal{P}_1, \mathcal{P}_2, \cdots, \mathcal{P}_K]^{T}$ are the powers allocated to each user, with $\sum_{k=1}^{K} \mathcal{P}_k= \mathcal{P}_{\text{tot}}$, where $\mathcal{P}_{\text{tot}}$ is the total transmit power.   The matrix $\mathbf{S} = \mathrm{diag}(s_1,s_2,\cdots,s_K) \in \mathbb{C}^{K \times K}$ contains the information-bearing signals chosen from a set of amplitude/phase symbols, and $\boldsymbol\varpi=[\varpi(1), \varpi(2), \cdots,\varpi(K) ]^{T} \in \mathbb{C}^{K \times 1}$ is the vector of precoder designed to eliminate interference. The noise vector $\mathbf{n}_B \in \mathbb{C}^{K \times 1}$ represents additive white Gaussian noise (AWGN) at all users and follows $\mathcal{CN}(0,\sigma_B^2 \mathbf{I}_K)$.

Focusing on the $k$th legitimate user, the received signal can be written as
\begin{equation}\label{receivedsignal}
y_{B}^{(k)} =
\underbrace{\mathbf{h}_{B}^{(k)\dagger} \boldsymbol{\Theta} \mathbf{G}_k \mathbf{x}_k}_{\text{Desired signal}}
+ \underbrace{\sum_{\substack{u=1 \\ u \neq k}}^{K} \mathbf{h}_{B}^{(k)\dagger} \boldsymbol{\Theta} \mathbf{G}_u \mathbf{x}_u}_{\text{Interference}}
+ n^{(k)}_B, \quad \forall k \in \mathcal{K},
\end{equation}
 where $\mathbf{h}_{B}^{(k)} \in \mathbb{C}^{M \times 1}$ is the RIS–user-$k$ channel vector, and $\mathbf{G}_k \in \mathbb{C}^{M \times n_t}$ denotes the AP–RIS subchannel corresponding to the $n_t$ antennas dedicated to user $k$. The effective transmitted signal is $\mathbf{x}_k = \sqrt{\mathcal{P}_k} \mathbf{w}_k  \varpi(k) s_k$, with $\varpi(k)$ being the precoder coefficient for user $k$, and $s_k$ is the normalized information symbol $\left(\mathbb{E}[|s_k|^2] = 1/K \right)$.

\section{Multi-User Data Detection with Precoder Design}

This section proposes a precoder design to eliminate MU interference by leveraging the CSI available at the transmitter. As previously introduced, the received signal at the legitimate user $k$ includes a summation term representing both the desired signal and interference from other users. To eliminate this interference, we design an optimal precoder, denoted as $\boldsymbol\varpi_{\text{opt}}$, which aims to achieve an error performance similar to that of interference-free transmission using a single-user maximum likelihood (ML) optimum detector. The precoder $\boldsymbol\varpi_{\text{opt}}$ is designed to orthogonalize the transmitted signals, thereby canceling out the interference terms.

For a given beamforming matrix $\mathbf{W}=\text{diag}\left( \mathbf{w}_1, \cdots, \mathbf{w}_K\right)$, phase shift matrix $\mathbf{\Theta}$, and power allocation vector $\boldsymbol{ \mathcal{P}},$ we can calculate the optimal precoder $\boldsymbol\varpi_{\text{opt}}$. The optimization of $\mathbf{W}$, $\mathbf{\Theta}$, and $\boldsymbol{ \mathcal{P}}$, aiming at maximizing the achievable secrecy rate, will be discussed in the next section. However, this current section focuses on eliminating MU interference via $\boldsymbol\varpi_{\text{opt}} $. The design of $\boldsymbol\varpi_{\text{opt}}$ begins by examining the general form of the received signal. To eliminate the interference terms, we require
\begin{equation}
\sum_{\substack{u=1 \\ u \neq k}}^{K} 
\mathbf{h}_{B}^{(k)\dagger} \mathbf{\Theta} \mathbf{G}_u \mathbf{x}_u 
= 0, \quad \forall k \in \mathcal{K}.
\end{equation}
This condition ensures that the received signal at user $k$ is free from interference caused by the signals intended for other users. To satisfy this condition, we construct the interference matrix $\mathbf{Q}\in \mathbb{C}^{K \times K}$ and the desired signal vector $\mathbf{q}\in \mathbb{C}^{K \times 1}$ as follows:
\begin{equation*}
\mathbf{Q} = \begin{bmatrix}
\sqrt{\mathcal{P}_1}\mathbf{h}_{B}^{(1)\dagger} \mathbf{\Theta} \mathbf{G}_1 \mathbf{w}_1 s_1 & \dots & \sqrt{\mathcal{P}_K}\mathbf{h}_{B}^{(1)\dagger} \mathbf{\Theta} \mathbf{G}_K \mathbf{w}_K s_K \\
\sqrt{\mathcal{P}_1}\mathbf{h}_{B}^{(2)\dagger} \mathbf{\Theta} \mathbf{G}_1 \mathbf{w}_1 s_1 & \dots & \sqrt{\mathcal{P}_K}\mathbf{h}_{B}^{(2)\dagger} \mathbf{\Theta} \mathbf{G}_K \mathbf{w}_K s_K \\
\vdots & \ddots & \vdots \\
\sqrt{\mathcal{P}_1}\mathbf{h}_{B}^{(K)\dagger} \mathbf{\Theta} \mathbf{G}_1 \mathbf{w}_1 s_1 & \dots & \sqrt{\mathcal{P}_K}\mathbf{h}_{B}^{(K)\dagger} \mathbf{\Theta} \mathbf{G}_K \mathbf{w}_K s_K   
\end{bmatrix},
\end{equation*}
$\mathbf{q} =[q_1,q_2,\cdots, q_K]^T$, where $q_k = \sqrt{\mathcal{P}_k}\mathbf{h}_{B}^{(k)\dagger} \mathbf{\Theta} \mathbf{G}_k \mathbf{w}_k s_k.$ The optimal precoder $\boldsymbol\varpi_{\text{opt}}$ is then derived by solving the following linear system:
\begin{equation}
\boldsymbol\varpi_{\text{opt}}= \mathbf{Q}^{-1} \mathbf{q}.
\end{equation}
By applying the optimal precoder $\boldsymbol\varpi_{\text{opt}}$, the received signal at user $k$ is interference-free and can be expressed as
\begin{equation} \label{Bobsignal}
y^{(k)}_{B} = \bar{\beta}\sqrt{\mathcal{P}_k}\mathbf{h}_{B}^{(k)\dagger} \mathbf{\Theta} \mathbf{G}_k \mathbf{w}_k s_k + n^{(k)}_B.
\end{equation}
This shows that with the precoder $\boldsymbol\varpi_{\text{opt}}$, the undesired interference term is eliminated, leaving only the desired signal and the noise term. To ensure that the average total transmitted power remains constant, we introduce a scaling factor $\bar{\beta}$, which is computed as
\begin{equation}\label{eq:beta}
\bar{\beta} =\mathbb{E}\left\{ \sqrt{\frac{K}{\operatorname{Tr}\!\left(\boldsymbol\varpi_{\text{opt}}\boldsymbol\varpi_{\text{opt}}^{\dagger}\right)}}\right\},
\end{equation} 
where $\mathbb{E}{\cdot}$ is the expectation operator. Note that by adapting the average power constraint at the transmitter, similar to the average constraint that is imposed on data symbols QAM/PSK in conventional communication systems, $\bar{\beta}$ becomes a constant. It can be computed by generating random realizations of $\boldsymbol{\Theta}$, $\mathbf{W}$, the fading channel, and signal constellations. It can then be passed on to each user. This would also contribute to a second-level security, since the eavesdropper does not have the $\bar{\beta}$ information to employ for detecting the user's data.

Finally, each legitimate user $k$ performs maximum likelihood (ML) detection on the interference-free signal:

\begin{equation}\label{det}
\hat{s}_{B}^{(k)} = \arg \min_{s_k\in \mathcal{S}} 
\left| y_{B}^{(k)} - \bar{\beta} \sqrt{\mathcal{P}_k}\mathbf{h}_{B}^{(k)\dagger} \boldsymbol{\Theta} \mathbf{G}_k  \mathbf{w}_{k} s_k \right|^2.
\end{equation}
This approach ensures that the system achieves optimal performance in terms of both signal quality and security, with the precoder $\boldsymbol\varpi_{\text{opt}}$ effectively eliminating MU interference.

\subsection{PLS with Precoded RIS-Aided CFmMIMO Network}

The first stage of the proposed PLS algorithm involves designing a spatial constellation. The optimal precoding vector, $\boldsymbol\varpi_{\text{opt}}$, is determined based on the CSI of the legitimate users, ensuring that legitimate users can reliably decode the transmitted signals while making it difficult for eavesdroppers to intercept and understand the communication. The RIS plays a crucial role in this process by assisting in beamforming, which involves calculating the optimal beamforming $\mathbf{W}_{\text{opt}}$, phase shift matrix $\mathbf{\Theta}_{\text{opt}}$ and power allocation $\boldsymbol{\mathcal{P}}_{\text{opt}}$ to align with the designed precoder.

If Eve receives the same signal that is intended for Bob, the received signal vector at Eve, denoted  $\mathbf{y}_E \in \mathbb{C}^{N_{E} \times 1}$, can be expressed as
\begin{equation}\label{yE}
\mathbf{y}_E = \bar{\beta} \mathbf{H}_E \mathbf{\Theta}_{\text{opt}} \mathbf{G} \mathbf{x} + \mathbf{n}_E,
\end{equation}
where $\mathbf{H}_E \in \mathbb{C}^{N_{E} \times M}$ represents the channel matrix between the RIS and Eve, and $\mathbf{G} \in \mathbb{C}^{M \times N_t}$ denotes the channel matrix from the AP to the RIS. The effective transmitted signal is given by $\mathbf{x} = \mathbf{W}_{\boldsymbol{\mathcal{P}}_{\text{opt}}}\mathbf{S} \boldsymbol\varpi_{\text{opt}} \in \mathbb{C}^{N_t \times 1}$, where  $\mathbf{S}$ is a diagonal matrix containing the information-bearing signals, and $\boldsymbol\varpi_{\text{opt}}$ is the optimal precoder vector. The noise vector $\mathbf{n}_E \sim \mathcal{CN}(0, \sigma_{E}^2 \mathbf{I}_{N_{E}})$ represents the AWGN at Eve.

When Eve attempts to intercept the signal intended for the legitimate user $k$, the received signal at Eve, denoted as $\mathbf{y}^{(k)}_{E}$, can be expressed as
\begin{equation} \label{Evesignal}
\mathbf{y}^{(k)}_{E} = \bar{\beta} \sqrt{\mathcal{P}_{k,\text{opt}}} \mathbf{H}_E \mathbf{\Theta}_{\text{opt}} \mathbf{G}_{k} \mathbf{w}_{k, \text{opt}} s_{k} + \mathbf{J}^{(k)}_{E} + \mathbf{n}^{(k)}_{E},
\end{equation}
where $\mathbf{J}^{(k)}_{E}$ represents the effective jamming signal experienced by Eve and from (\ref{yE}) and (\ref{Evesignal}), it can be expressed as
\begin{equation} \label{eq:jamming}
\mathbf{J}^{(k)}_{E}\! = \!\bar{\beta}\mathbf{H}_E \mathbf{\Theta}_{\text{opt}} \left( \mathbf{G} \mathbf{W}_{\boldsymbol{\mathcal{P}},\text{opt}}  \mathbf{S} \boldsymbol\varpi_{\text{opt}} \!-\! \sqrt{\mathcal{P}_{k,\text{opt}}} \mathbf{G}_{k} \mathbf{w}_{k,\text{opt} } s_{k}\right).
\end{equation}
This effective jamming signal arises inherently from the design of the optimal precoder $\boldsymbol\varpi_{\text{opt}}$, which is optimized based on the legitimate users’ CSI. The RIS combined with $\boldsymbol\varpi_{\text{opt}}$ differs from schemes where a separate AN signal is intentionally transmitted, which typically reduces power efficiency. 
Instead, it enhances constructive interference toward legitimate users while simultaneously inducing destructive interference at Eve’s receiver. Consequently, the signal received by Eve, $\mathbf{y}^{(k)}_{E}$, is heavily distorted, preventing accurate signal recovery and significantly degrading the BER performance, thereby ensuring robust physical layer security. $\mathbf{J}^{(k)}_{E}$ is approximately Gaussian vector with zero mean and variance $\sigma^{2}_{\mathbf{J}^{(k)}_{E}}$. The following lemma derives the exact variance of the jamming signal.  

\textit{Lemma 1:} The variance of the jamming signal is given as  
\begin{align}
\sigma^{2}_{\mathbf{J}^{(k)}_{E}} 
&= \mathbb{E}_{\mathbf{H}_{E},\mathbf{G}} \big\{ \mathbf{J}^{(k)\dagger}_{E} \mathbf{J}^{(k)}_{E} \big\} \\
&= M N_E \bar{\beta}^{2} \left( \mathcal{P}_{\mathrm{tot}}\left(\bar{\beta}^{-2} + K^{-1}\right) + 2\mathbb{R}\{\varpi_{opt} (k)\}\right),\nonumber
\end{align}
where $\mathcal{P}_{\mathrm{tot}}$ is the total transmit power and $\bar{\beta}$ is the power normalizing constant defined by (\ref{eq:beta}).

\textit{Proof:} The proof is given in Appendix A.

Similar to (\ref{det}), the eavesdropper’s data detection with ML criterion is expressed as
\begin{equation}
\hat{s}_{E}^{(k)} = \arg\min_{s_k \in \mathcal{S}}
\left\| \mathbf{y}_{E}^{(k)} - \bar{\beta} \sqrt{\mathcal{P}_{k,\text{opt}}} \mathbf{H}_E \boldsymbol{\Theta}_{\text{opt}} \mathbf{G}_k \mathbf{w}_{k,\text{opt}}  s_k \right\|^2.
\end{equation}
This process ensures that even with optimal detection strategies, Eve’s ability to intercept and decode the signal successfully is significantly impaired, reinforcing the security provided by the RIS-aided CFmMIMO network.

\section{Optimization Problem}
In this section, we formulate secrecy sum–rate maximization by jointly optimizing the transmit beamformer vectors $\{\mathbf{w}_k\}$, the user power coefficients $\{\mathcal{P}_k\}$, and the RIS phase–shift matrix $\boldsymbol{\Theta}\!=\!\mathrm{diag}(e^{j\theta_1},\ldots,e^{j\theta_M})$. The achievable secrecy rate for user $k$ is defined as
\begin{equation}
    R_s^{(k)} = \big[ R_{B}^{(k)} - R_{E}^{(k)} \big]^+, \label{eq:Rs_k_final}
\end{equation}
where $[x]^{+} = \max\{0,x\}$ and $R_{B}^{(k)}$ denotes the achievable rate at the legitimate user and $R_{E}^{(k)}$ represents the rate at Eve attempting to decode the message of user $k$.

From the received signal of user $k$ in (\ref{Bobsignal}), noting that $\mathbb{E}\{|s_k|^2\} = 1/K$, the achievable rate is expressed as
\begin{equation}\label{eq:Rbk_final}
R_{B}^{(k)}= \log_{2}\!\left( 1 + 
\frac{\bar{\beta}^2\mathcal{ P}_k\big|\mathbf{h}_{B}^{(k)\dagger} \boldsymbol{\Theta} \mathbf{G}_k \mathbf{w}_k\big|^2}{K \sigma_B^2} \right),
\end{equation}
where $\mathbf{h}_{B}^{(k)}\in\mathbb{C}^{M\times 1}$ denotes the RIS–user channel and $\sigma_B^2$ is the noise variance at the legitimate receiver.

Similarly, from Eve’s received signal in (\ref{Evesignal}), the achievable rate is initially written as
\begin{equation} 
    R_{E}^{(k)}
= \log_2 \left( 1 
     + \frac{\bar{\beta}^2\mathcal{ P}_k\|\mathbf{H}_E \boldsymbol{\Theta} \mathbf{G}_k \mathbf{w}_k \|^2}
     {K (\sigma_E^2 + \sigma^{2}_{\mathbf{J}^{(k)}_{E}})}
 \right).
    \label{eq:Rek_logdet}
\end{equation}

Letting
\begin{equation}\label{eq:identity}
\Psi_{B}^{(k)} = \frac{\bar{\beta} \mathbf{h}_{B}^{(k)\dagger} \boldsymbol{\Theta} \mathbf{G}_k}{\sqrt{K}}, 
\qquad 
\Psi_{E}^{(k)} = \frac{\bar{\beta} \mathbf{H}_E \boldsymbol{\Theta} \mathbf{G}_k}{\sqrt{K}},
\end{equation}
the resulting achievable secrecy rate of user $k$ can be rewritten as

\begin{equation}\label{RISrate}
 R_s^{(k)}\! =\!\!\!\left[\log_2 \!\left(\!\!\! 1 +\! \frac{\mathcal{P}_k |\Psi_{B}^{(k)} \mathbf{w}_k|^2}{\sigma_B^2}\! \right) 
\!\!- \log_2 \!\!\left(\!\!\! 1 + \!\frac{\mathcal{P}_k \|\Psi_{E}^{(k)} \mathbf{w}_k\|^2}{\sigma_E^2 + \sigma^{2}_{J^{(k)}_{E}}} \!\right)\!\!\right] ^+ .
\end{equation}
Here, $\mathcal{P}_k$ represents the total transmit power available for the $k$th beamforming vector. 

We note that dropping $[\,\cdot\,]^+$ in (\ref{RISrate}) has no impact on the optimization \cite{Shafiee2007}. Consequently, the overall secrecy sum–rate is given by
\begin{equation} \label{eq:sum_opt}
R_s = \sum_{k=1}^{K}  R_s^{(k)}.
\end{equation}

The secrecy sum–rate maximization problem can now be formulated as
\begin{equation} \label{eq:secrecyrate}
\hspace{-5cm}\max_{\{\mathcal{P}_k\},\{\mathbf{w}_k\}, \, \boldsymbol{\Theta}} \; R_s \nonumber
\end{equation}
\begin{subequations}
\begin{equation}
\hspace{-5cm}\text{s.t.}\quad \|\mathbf{w}_k\|^2 = 1,
\end{equation}
\begin{equation}  \label{19b}
\sum_{k=1}^K \mathcal{P}_k= \mathcal{P}_{\text{tot}}, \quad\mathcal{P}_k \geq 0,\quad 
k = 1, 2, \ldots, K.
\end{equation}
\begin{equation}
\hspace{-2cm}\boldsymbol{\Theta} = \mathrm{diag}\left(e^{j\theta_1}, e^{j\theta_2}, \cdots, e^{j\theta_M}\right)
\end{equation}
\begin{equation} \label{19d}
\hspace{-2cm}|e^{j\theta_m}| = 1, \quad \theta_m \in [0, 2\pi), \; \forall m.
\end{equation}
\end{subequations}
Here, the constraint (\ref{19b}) ensures that the total power consumed by all beamforming vectors does not exceed $\mathcal{P}_{\text{tot}}=\sum_{k=1}^K\mathcal{P}_k$. The constraint in (\ref{19d}) enforces the unit modulus property of each RIS element.

To efficiently solve the non-convex optimization problem in \eqref{eq:secrecyrate}, we employ an alternating optimization framework that iteratively updates the beamforming vectors, the user power allocation, and the RIS phase-shift matrix. The transmitter utilizes a CSI-based precoder to eliminate multi-user interference, enabling interference-free signal reception for legitimate users. Based on this precoding structure, the achievable secrecy rate can be directly evaluated as given in \eqref{RISrate}. In each iteration, the beamforming vectors are optimized for the current RIS configuration and power allocation, followed by refinement of the user power coefficients under the total power constraint. Finally, RIS phase-shift optimization is formulated under the RMO framework and solved using a Riemannian Conjugate Gradient (RCG) method on the complex unit-modulus manifold. The overall optimization framework, which integrates beamforming, power allocation, and RIS phase shift, is summarized in Algorithm \ref{algorithm}, achieving stable convergence to a locally optimal solution for the secrecy sum rate.
\begin{algorithm}
	\vspace{2mm}
	\caption{Alternating Optimization Algorithm for Achievable Secrecy Sum Rate}
	\label{algorithm}
1.	{\bf Input:} $K$, initial power allocation $\mathcal{P}_{0}=\mathcal{P}_{\text{tot}}*[1/K,1/K, \cdots, 1/K]$,  RIS phase shifts $\mathbf{v}_0$ ($\bf{\Theta_{0}})$, and convergence tolerance $\epsilon$   \\
2.	{\bf Output:} optimized $R_s$\\
3.	{\bf repeat}\\
4. Solve optimization algorithm (\ref{eq:opt_problem_2}) to obtain $\mathbf{W}_{0}=\{\mathbf{w}_{0,k}\}_{k=1}^{K}$\\
5.  Compute $R^{(\text{prev)}}_s= R_s\left(\bf{\Theta}_{0}, \mathcal{P}_0,\mathbf{W}_{0}\right)$ according to (\ref{RISrate}). \\
6. Update the counter: $i \leftarrow  i+1$\\
7. Solve optimization algorithm (\ref{RISphase}) to obtain 
$\left\{\mathcal{P},\bf{\Theta}\right\}$\\ 
8. Solve optimization algorithm (\ref{eq:opt_problem_2}) to obtain $\mathbf{W}$\\
9.  Compute $R_s= R_s\left(\bf{\Theta}, \mathcal{P},\mathbf{W}\right)$\\
10. {\bf until} $ \|R_s-R^{\text{(prev)}}_{s} \|<\epsilon $\\
11.  {\bf return} $R_{s}^\text{opt}$ 
\end{algorithm}  

\subsection{Power Allocation Optimization}
For given $\bf{\Theta}$ and $\mathbf{W}$, the goal is to maximize the system's achievable secrecy sum rate by assigning optimal powers to each user's beamforming vector $\mathbf{w}_k$.   
Note that the powers $\mathcal{P}=\left[\mathcal{P}_1, \mathcal{P}_2, \cdots,\mathcal{P}_{K}\right]^{T}$, with $\sum_{k=1}^{K} \mathcal{P}_k= \mathcal{P}_{\text{tot}}$,  allocated to beamforming vectors to each user are adjusted to maximize the system's achievable secrecy sum rate according to the strength of the eavesdropper's distance from the legitimate users. Hence, by normalizing the allocated powers as
\begin{equation} \label{eq:norm}
	\bar{\mathcal{P}}_k = \frac{\mathcal{P}_k}{\mathcal{P}_{\text{tot}}}, 
	\qquad \sum_{k=1}^{K}\bar{\mathcal{P}}_k = 1,
\end{equation}
the achievable secrecy rate,  $ R_s^{(k)}$, for the $k$th user for $k=1,2,\cdots, K,$ with power allocation is given by
 \begin{equation}
 	 R_s^{(k)}= \left[ \log_2\left(1+ \bar{\mathcal{P}}_k a_k\right)- \log_2\left(1+\bar{\mathcal{P}}_k b_k\right) \right]^{+}\!, k=1,2,\cdots,K
 \end{equation}
where from Eq. (\ref{RISrate}), $a_k$ and $b_k$  are defined  as
 \begin{equation}
 	a_k=\frac{ \mathcal{P}_{\text{tot}}| \Psi^{(k)}_B \mathbf{w}_k |^2}{\sigma^{2}_B}, \hspace{3mm} 	b_k = \frac{ \mathcal{P}_{\text{tot}} \left\| \Psi^{(k)}_E \mathbf{w}_k \right\|^2 }{ \sigma^{2}_{E} + \sigma^{2}_{J^{(k)}_E} }.
 \end{equation} 
 
We can now characterize the boundary of the achievable secrecy regions of the proposed scheme, for a given $\bf{\Theta}$ and $\bf{W}$,  by solving the following optimization problem. 
\begin{eqnarray}\label{opt}
	&&\hspace*{-0.5cm} \max_{\bar{\mathcal{P}}} ~ R_s(\bar{\mathcal{P}})  \nonumber \\
	&\mbox{s.t.}&    \sum_{k=1}^{K} \bar{\mathcal{P}}_k = 1, ~~ 0\leq \bar{\mathcal{P}}_k  \leq 1 .
\end{eqnarray}

Depending on the values of $a_k$ and $b_k$, the above problem can then be solved analytically, leading to a unique global optimal solution for the power allocation problem as described step by step below.
\subsubsection{Problem and basic simplifications}
Define the single-user objective (before the $[\cdot]_+$ operator) as
 \begin{equation}	
 	\begin{array}{l}
 			f_k(\bar{\mathcal{P}}_k) \triangleq \log_2(1 + a_k \bar{\mathcal{P}}_k) - \log_2(1 + b_k \bar{\mathcal{P}}_k).
 		\label{eq:fk}
 		
 	\end{array}	
 \end{equation} 
 Its first derivative is
 \begin{equation}
 	\begin{array}{l}
 	f_k'(\mathcal{\bar{\mathcal{P}}}_k) 
 	= \frac{1}{\ln 2} \!\left( \frac{a_k}{1 + a_k \bar{\mathcal{P}}_k} - \frac{b_k}{1 + b_k \bar{\mathcal{P}}_k} \right)    \vspace{1mm}\\
 	
 	\;\;\;\;\;\;\;\;\;\;\;\; = \frac{a_k - b_k}{\ln 2 \, (1 + a_k \bar{\mathcal{P}}_k)(1 + b_k \bar{\mathcal{P}}_k)}.
 	
 	\label{eq:fkprime}
 \end{array}
 \end{equation}
 Hence:
 \begin{itemize}
 	\item If $a_k \le b_k$, then $f_k'(\bar{\mathcal{P}}_k) \le 0$ and $f_k(0) = 0$. Under the sum-power constraint, allocating any positive power to such a user decreases the total objective (it ``steals'' power from others while being non-increasing itself). Therefore, the optimum is $\bar{\mathcal{P}}_k^\star = 0$ for all $k$ with $a_k \le b_k$.
 	\item If $a_k > b_k$, then $f_k'(\bar{\mathcal{P}}_k) > 0$ for all $\bar{\mathcal{P}}_k \ge 0$ and strictly decreases with $\bar{\mathcal{P}}_k$. These users form the \emph{active set}
 	\(
 	\mathcal{A} \triangleq \{ k : a_k > b_k \}.
 	\)
 \end{itemize}
 We only need to allocate power over $\mathcal{A}$; all other users receive zero power.
 \subsubsection{Concavity and uniqueness on the active set}
 For $a_k > b_k$, the second derivative is
 \begin{equation}
 	\begin{array}{l}		
 	
 	f_k''(\bar{\mathcal{P}}_k)
 	= \frac{1}{\ln 2} \!\left( -\frac{a_k^2}{(1 + a_k \bar{\mathcal{P}}_k)^2} + \frac{b_k^2}{(1 + b_k \bar{\mathcal{P}}_k)^2} \right) < 0,
 	\qquad \forall \bar{\mathcal{P}}_k \ge 0,
 	\label{eq:fksecond}
 	\end{array}	
 \end{equation}
 since $x \mapsto \frac{x^2}{(1+x\bar{\mathcal{P}}_k)^2}$ is increasing for $x>0$ and $a_k>b_k$. Therefore, over the feasible set
 \begin{equation}
 	\begin{array}{l}
 		\max\limits_{\{\bar{\mathcal{P}}_k\}} \ \sum_{k \in \mathcal{A}} f_k(\bar{\mathcal{P}}_k)   \vspace{1mm}\\

 		 \text{s.t.} \quad \sum_{k \in \mathcal{A}} \bar{\mathcal{P}}_k = 1, \quad 0 \le \bar{\mathcal{P}}_k \le 1,
 		\label{eq:active-problem}
 		
 	\end{array}	 	
 \end{equation}
the objective is strictly concave, and the constraints are convex, so the problem is a convex optimization with a \emph{unique} global maximizer characterized by the KKT conditions.
 
 \subsubsection{KKT conditions}
 Consider the Lagrangian (restricted to $\mathcal{A}$):
 \begin{equation}
 	\begin{array}{l}
 	\!\!\!\!\!\!\!\!\!\!\!\!	\mathcal{L}(\bar{\mathcal{P}},\mu,\nu,\tau)
 		= \sum\limits_{k \in \mathcal{A}} f_k(\bar{\mathcal{P}}_k)
 		- \mu \!\left( \sum\limits_{k \in \mathcal{A}} \bar{\mathcal{P}}_k - 1 \right)   \vspace{1mm}\\
 		
 	\;\;\;\;\;\;\;\;\;\;\;\;\;\;\;\;	
 	
 	    + \sum\limits_{k \in \mathcal{A}} \nu_k \bar{\mathcal{P}}_k
 		+ \sum\limits_{k \in \mathcal{A}} \tau_k (1 - \bar{\mathcal{P}}_k),
 		\label{eq:lagrangian}
 		
 	\end{array}	 	
 \end{equation}
 where $\mu \in \mathbb{R}$ is the multiplier for the equality constraint and $\nu_k, \tau_k \ge 0$ correspond to the lower/upper bounds. The KKT conditions are:
  
  \text{Primal feasibility:}  
        \begin{equation}	
        	\begin{array}{l}
        		\sum\limits_{k \in \mathcal{A}} \bar{\mathcal{P}}_k = 1, \quad 0 \le \bar{\mathcal{P}}_k \le 1.

        		 \label{eq:primal-feas}          		
        	\end{array}	
        \end{equation}

 	 \text{Stationarity:} 
 	 \begin{equation}	
 	 	\begin{array}{l}
 	 		f_k'(\bar{\mathcal{P}}_k) - \mu + \nu_k - \tau_k = 0, \quad \forall k \in \mathcal{A}.   
 	 		
 	 		\label{eq:stationarity}
 	 	\end{array}	
 	 \end{equation}

 	 \text{Complementary slackness:}  
 	 \begin{equation}	
 	 	\begin{array}{l}
 	 		 \nu_k \bar{\mathcal{P}}_k = 0, \quad \tau_k (1 - \bar{\mathcal{P}}_k) = 0,\quad \forall k \in \mathcal{A}. 
 	 		 
 	 		 \label{eq:cs} 	 		
 	 	\end{array}	
 	 \end{equation}
 	 
 \subsubsection{Closed-form interior solution ($0 < \bar{\mathcal{P}}_k < 1$)}
 If, at the optimum, a user $k \in \mathcal{A}$ satisfies $0 < \bar{\mathcal{P}}_k^\star < 1$, then $\nu_k = \tau_k = 0$ and
 \begin{equation}
 	\begin{array}{l}
 		f_k'(\bar{\mathcal{P}}_k^\star) = \mu.
 		
 		\label{eq:interior-stationarity}
 	\end{array}	
 \end{equation}
 Let $c \triangleq \mu \ln 2 > 0$. Using \eqref{eq:fkprime}, the stationarity condition becomes
 \begin{equation}
 	\begin{array}{l}
 		\frac{a_k - b_k}{(1 + a_k \bar{\mathcal{P}}_k^\star)(1 + b_k \bar{\mathcal{P}}_k^\star)} = c
 		\vspace{1mm}\\

 		 \Longleftrightarrow  
 		(1 + a_k \bar{\mathcal{P}}_k)(1 + b_k \bar{\mathcal{P}}_k) = \frac{a_k - b_k}{c}.
 		\label{eq:quad-setup}
 		
 	\end{array}	 	
 \end{equation}
 This yields a quadratic equation in $P$:
 \begin{equation}
 	\begin{array}{l}
 		a_k b_k \bar{\mathcal{P}}_k^2 + (a_k + b_k) \bar{\mathcal{P}}_k + 1 - \frac{a_k - b_k}{c} = 0.
 		\label{eq:quad}
 		
 	\end{array}	 	
 \end{equation}
 The discriminant is
 \begin{equation}
 	\begin{array}{l}
 	\!\!\!\!\!\!\!\!\!\!	\Delta_k
 		= (a_k + b_k)^2 - 4 a_k b_k \! ( 1 - \frac{a_k - b_k}{c} )   \vspace{1mm}\\

 		= (a_k - b_k)^2 + \frac{4 a_k b_k (a_k - b_k)}{c}  \vspace{1mm}\\
 		
 		\; > \; (a_k - b_k)^2.
 		\label{eq:discriminant}
 		
 	\end{array}	 	
 \end{equation}
 The only nonnegative root (the other is negative) is
 \begin{equation}
 	\begin{array}{l}
 	\!\!\!\!\!\!\!\!\!\!\!\!\!\!\!\!\!\!\!\!\!	\bar{\mathcal{P}}_k^{\mathrm{int}}(c)
 		= \frac{ - (a_k + b_k) + \sqrt{\Delta_k} }{ 2 a_k b_k }  \vspace{1mm}\\

 		= \frac{ - (a_k + b_k) + \sqrt{ (a_k - b_k)^2 + \frac{4 a_k b_k (a_k - b_k)}{c} } }{ 2 a_k b_k }.
 		\label{eq:Pk-int-c}
 		
 	\end{array}	 	
 \end{equation}
 Equivalently, in terms of $\mu$:
 \begin{equation}
 	\begin{array}{l}
 	\!\!\!\!\!\!\bar{\mathcal{P}}_k^{\mathrm{int}}(\mu)
 		= \frac{ - (a_k + b_k) + \sqrt{ (a_k - b_k)^2 + \frac{4 a_k b_k (a_k - b_k)}{ \mu \ln 2 } } }{ 2 a_k b_k },      
 		  a_k > b_k.
 		\label{eq:Pk-int-mu}
 		
 	\end{array}	 	
 \end{equation}
 One can verify that $\bar{\mathcal{P}}_k^{\mathrm{int}}(\mu)$ is strictly decreasing in $\mu$.
 
 \subsubsection{Boundary cases and projection}
 If the interior formula \eqref{eq:Pk-int-mu} yields a value outside $[0,1]$, the bound(s) must be active by complementary slackness:
 \[
 \bar{\mathcal{P}}_k^\star = 0 \ \text{if}\ \bar{\mathcal{P}}_k^{\mathrm{int}}(\mu) \le 0,
 \qquad
 \bar{\mathcal{P}}_k^\star = 1 \ \text{if}\ \bar{\mathcal{P}}_k^{\mathrm{int}}(\mu) \ge 1.
 \]
 Hence, the single-user solution can be written compactly via projection:
 \begin{equation}
 	\begin{array}{l}
 		\bar{\mathcal{P}}_k^\star(\mu) 
 		= \big[\, \bar{\mathcal{P}}_k^{\mathrm{int}}(\mu) \,\big]_0^1
 		\triangleq \min\{1,\ \max\{0,\ \bar{\mathcal{P}}_k^{\mathrm{int}}(\mu)\}\},
 		  k \in \mathcal{A};   \\
 		 
 		\bar{\mathcal{P}}_k^\star = 0, \ k \notin \mathcal{A}.
 		\label{eq:projection}
 		
 	\end{array}	 	
 \end{equation}
 
 \subsubsection{Determining the water level $\mu$}
 Define
 \begin{equation}
 	\begin{array}{l}
 		\phi(\mu) \triangleq \sum_{k \in \mathcal{A}} \bar{\mathcal{P}}_k^\star(\mu).
 		\label{eq:phi}
 		
 	\end{array}	 	
 \end{equation}
 Since each $\bar{\mathcal{P}}_k^\star(\mu)$ is nonincreasing in $\mu$, $\phi(\mu)$ is strictly decreasing. Moreover,   
 \begin{equation}	
 	\begin{array}{l}
 		\lim_{\mu \downarrow 0} \phi(\mu) \ge 1,   \\
 		 
 		\lim_{\mu \uparrow \infty} \phi(\mu) = 0.
 		
 	\end{array}	
 \end{equation}  
 Therefore, there exists a unique $\mu^\star > 0$ such that
 \begin{equation}	
 	\begin{array}{l}
 		\sum_{k \in \mathcal{A}} \bar{\mathcal{P}}_k^\star(\mu^\star) = 1.
 		\label{eq:power-balance}
 		
 	\end{array}	
 \end{equation}  
 The scalar $\mu^\star$ (the \emph{water level}) can be efficiently found by a 1D search (e.g., bisection or Newton), after which $\bar{\mathcal{P}}^\star$ follows from \eqref{eq:projection}. This yields a \emph{semi-analytical} solution: the per-user powers admit a closed form given $\mu$, while $\mu^\star$ is determined as the unique root of the one-dimensional equation \eqref{eq:power-balance}.

 \subsubsection{Degenerate and special cases}
 \begin{itemize}
 	\item \textbf{Inactive users:} If $a_k \le b_k$, then $\bar{\mathcal{P}}_k^\star = 0$, which is consistent with the $[\cdot]_+$ structure.
 	\item \textbf{Tie case:} If $a_k = b_k$, then $f_k \equiv 0$ and allocating power is pointless; we may set $\bar{\mathcal{P}}_k^\star = 0$.
 	\item \textbf{No eavesdropper ($b_k = 0$):} \eqref{eq:Pk-int-mu} reduces to the classical water-filling form,
 	\(
 	\bar{\mathcal{P}}_k^{\mathrm{int}}(\mu) = \frac{1}{\mu \ln 2} - \frac{1}{a_k},
 	\)
 	followed by projection onto $[0,1]$.
 	\item \textbf{Many active upper bounds:} If several users saturate at $\bar{\mathcal{P}}_k^\star = 1$, the sum constraint $\sum_k \bar{\mathcal{P}}_k^\star = 1$ automatically forces the remainder to $0$ (or only one user equals $1$), maintaining consistency.
 \end{itemize}

\subsubsection{Final solution and complexity}
 The globally optimal solution is:
 \begin{enumerate}
 	\item Form the active set $\mathcal{A} = \{ k : a_k > b_k \}$; set $\bar{\mathcal{P}}_k^\star = 0$ for $k \notin \mathcal{A}$.
 	\item For $k \in \mathcal{A}$,
 	\begin{equation}	
 		\begin{array}{l}
 		\!\!\!\!\!\!\!\!\!\!\!\!	\bar{\mathcal{P}}_k^\star(\mu)
 			= \left[
 			\frac{ - (a_k + b_k) + \sqrt{ (a_k - b_k)^2 + \dfrac{4 a_k b_k (a_k - b_k)}{ \mu \ln 2 } } }{ 2 a_k b_k }
 			\right]_0^1 .
 			
 			\label{eq:final-Pk} 			
 		\end{array}	
 	\end{equation} 	
 	\item Find the unique $\mu^\star > 0$ satisfying $\sum_{k \in \mathcal{A}} \bar{\mathcal{P}}_k^\star(\mu^\star) = 1$ via a 1D search.
 	\item Substitute $\mu^\star$ into \eqref{eq:final-Pk} to obtain $\bar{\mathcal{P}}^\star$.
 \end{enumerate}
 Each evaluation of $\phi(\mu)$ costs $O(|\mathcal{A}|)$, and a bisection method converges in $O(\log(1/\varepsilon))$ iterations to accuracy $\varepsilon$, yielding an overall light-weight and stable procedure.

\subsection{Beamforming Optimization}\label{w_opt}
The joint optimization problem formulated in \eqref{eq:secrecyrate} is non-convex due to the coupling between the beamforming vectors and the RIS phase shifts. To facilitate its solution, we first focus on the beamformer design. Notably, each term in \eqref{eq:secrecyrate} depends only on the variables of user $k$ and can thus be optimized separately. Consequently, for a given power allocation $\mathcal{P}_k=\bar{\mathcal{P}}_k \mathcal{P}_{\text{tot}}$ and the RIS phase–shift matrix $\boldsymbol{\Theta}$, using the identities in (\ref{eq:identity}), the secrecy rate of user k can be expressed as
\begin{subequations}\label{eq:opt_problem_2}
\begin{align}
\max_{\mathbf{w}_k} \quad &
\log_2\!\left(
\frac{1 + \tfrac{\mathcal{P}_k}{\sigma_{B}^2}\,|\Psi_{B}^{(k)}\mathbf{w}_k|^2}
     {1 + \tfrac{\mathcal{P}_k}{\sigma_{E}^2+\sigma^{2}_{\mathbf{J}^{(k)}_{E}}}\,\|\Psi_{E}^{(k)}\mathbf{w}_k\|^2}
\right) \label{eq:opt_problem_2a} \\
\text{s.t.} \quad 
& \|\mathbf{w}_k\|^2 \leq 1 .\label{eq:opt_problem_2b}
\end{align}
\end{subequations}
Defining,
\begin{equation}
X_{B}^{(k)} = \mathbf{I}_{n_t} + \frac{\mathcal{P}_k}{\sigma_B^2} 
\Psi_{B}^{(k)\dagger} \Psi_{B}^{(k)},
\end{equation}
\begin{equation}
X_{E}^{(k)} = \mathbf{I}_{n_t} + \frac{\mathcal{P}_k}{\sigma_E^2 + \sigma_{\mathbf{J}_{E}^{(k)}}^{2}} 
\Psi_{E}^{(k)\dagger} \Psi_{E}^{(k)},
\end{equation}
the optimal beamformer that maximizes the secrecy rate of user $k$ is given by \cite{Shafiee2007}.
\begin{equation} \label{eq:opt_w}
\mathbf{w}_k^{\mathrm{opt}} = \sqrt{\mathcal{P}_k} \, \mathbf{u}_{\max}\left(\left(X_{E}^{(k)}\right)^{-1} X_{B}^{(k)}\right), \quad k = 1,2,\cdots,K
\end{equation}
with $\|\mathbf{w}^{\mathrm{opt}}_k\|^2 = 1$.  $\mathbf{u}_{\max}(\cdot)$ denotes the eigenvector corresponding to the 
maximum eigenvalue (i.e., the dominant eigenvector).

\subsection{RIS Phase–Shift Optimization}\label{theta_opt}
Finally, for the optimized beamformers $\{\mathbf{w}_k^{\mathrm{opt}}\}$ the achievable secrecy sum rate $R_s$ in (\ref{eq:sum_opt}) is further influenced by the RIS phase–shift matrix $\boldsymbol{\Theta} = \mathrm{diag}(e^{j\theta_1}, e^{j\theta_2}, \ldots, e^{j\theta_M})$.

Substituting (\ref{eq:opt_w}) into (\ref{eq:opt_problem_2}), the secrecy rate of user $k$ becomes
\begin{equation}
R_k^{\mathrm{opt}} 
= \log_2\!\left(
\frac{
1 + \dfrac{\mathcal{P}_k}{\sigma_B^2}
\big| \Psi_{B}^{(k)} \mathbf{w}_k^{\mathrm{opt}} \big|^2
}{
1 + \dfrac{\mathcal{P}_k}{\sigma_E^2 + \sigma^{2}_{\mathbf{J}^{(k)}_{E}}}
\big\| \Psi_{E}^{(k)} \mathbf{w}_k^{\mathrm{opt}} \big\|^2
}
\right).
\end{equation}
Accordingly, the secrecy sum rate is expressed as
\begin{equation} \label{eq:opt_sum_rate}
R_s^{\mathrm{opt}} = \sum_{k=1}^{K} R_k^{\mathrm{opt}}.
\end{equation}
Thus, the RIS phase–shift optimization problem is formulated as
\begin{equation} \label{RISphase}
\begin{aligned}
\max_{\boldsymbol{\Theta}} \quad & R_s^{\mathrm{opt}}(\boldsymbol{\Theta}) \\
\text{s.t.} \quad & |e^{j\theta_m}| = 1, \quad m = 1,\ldots,M.
\end{aligned}
\end{equation}

This optimization problem is non–convex due to the unit–modulus constraints, making it challenging to obtain a global optimum.

Let $\mathbf{v} = [e^{j\theta_1}, e^{j\theta_2}, \ldots, e^{j\theta_M}]^\dagger$, 
$\mathbf{a}_k = \mathbf{G}_k \mathbf{w}_k^{\mathrm{opt}}$, 
$\mathbf{b}_k = \mathbf{h}_{B}^{(k)}$, and 
$\mathbf{C}_k = \mathbf{H}_E$. 
Then, the rate ratio $F_k(\mathbf{v})$ for user $k$ can be written as
\begin{equation}
F_k(\mathbf{v}) = \frac{1 + A_k |\mathbf{v}^\dagger \mathbf{u}_k|^2}{1 + B_k \mathbf{v}^\dagger \mathbf{D}_k \mathbf{v}},
\end{equation}
where $A_k = \mathcal{P}_k \,\bar{\beta}^2/({K\,\sigma_B^2})$, $B_k = \mathcal{P}_k \bar{\beta}^2/K(\sigma_E^2 + \sigma^{2}_{\mathbf{J}^{(k)}_{E}})$, 
$\mathbf{u}_k = \mathrm{diag}(\mathbf{b}_k^\dagger)\mathbf{a}_k$, 
and $\mathbf{D}_k = \mathrm{diag}(\mathbf{a}_k^\dagger)\mathbf{C}_k^\dagger \mathbf{C}_k \mathrm{diag}(\mathbf{a}_k)$.

Next, using the multidimensional complex quadratic transform~\cite{Shen2018}, 
we can convert the fractional–form phase–shift optimization problem (\ref{eq:opt_sum_rate}) into the following form:
\begin{equation} \label{eq:quadratic}
\begin{aligned}
\max_{\mathbf{v},\boldsymbol{\gamma}} \quad \!\!\!&\!\!\!
\sum_{k=1}^{K} 
\Big[ 2 \sqrt{1 + A_k |\mathbf{v}^\dagger \mathbf{u}_k|^2} \cdot \mathfrak{Re}\{\gamma_k^{\ast}\} 
- |\gamma_k|^2 \big(1 + B_k \mathbf{v}^\dagger \mathbf{D}_k \mathbf{v}\big)\! \Big] \\
\text{s.t.} \quad & |\mathbf{v}_m|=1, \quad m=1,\ldots,M,
\end{aligned}
\end{equation}
where $\gamma_k \in \mathbb{C}$ is an auxiliary variable, $(\cdot)^\ast$ denotes the complex conjugate, and $\mathfrak{Re}\{\cdot\}$ represents the real part.

For each $k$, the above objective function is a concave quadratic function with respect to $\gamma_k$. By setting its derivative to zero, a closed-form solution can be obtained:
\begin{equation} \label{eq:gamma_opt}
\gamma_k^{\mathrm{opt}} = \frac{\sqrt{1 + A_k |\mathbf{v}^\dagger \mathbf{u}_k|^2}}{1 + B_k \mathbf{v}^\dagger \mathbf{D}_k \mathbf{v}}, \quad \forall k.
\end{equation}

For fixed $\gamma_k$, problem (\ref{eq:quadratic}) can be converted into:
\begin{equation} \label{eq:phase_opt}
\begin{aligned}
\max_{\mathbf{v}} \quad & \sum_{k=1}^{K} 
\Big[ 2 \sqrt{1 + A_k |\mathbf{v}^\dagger \mathbf{u}_k|^2} \cdot \mathfrak{Re}\{\bar{\gamma}_k^{\ast}\} 
- |\bar{\gamma}_k|^2 (1 + B_k \mathbf{v}^\dagger \mathbf{D}_k \mathbf{v}) \Big] \\
\text{s.t.} \quad & |\mathbf{v}_m| = 1, \quad m = 1,\ldots,M,
\end{aligned}
\end{equation}

However, the objective function of (\ref{eq:phase_opt}) remains a non-convex function and cannot be solved directly. To facilitate the solution of subsequent problems, we use the first-order Taylor expansion to further handle the non-convex part of the objective function. Therefore, problem (\ref{eq:phase_opt}) can be rewritten as
\begin{equation} \label{eq:re_phase_opt}
\begin{aligned}
\min_{\mathbf{v}} \quad & \sum_{k=1}^{K} 
\Big[ |\bar{\gamma}_k|^2 B_k \mathbf{v}^\dagger \mathbf{D}_k \mathbf{v} - 2 g^{(t)} \cdot \mathfrak{Re}\{\bar{\gamma}_k^{\ast}\} \Big] \\
\text{s.t.} \quad & |\mathbf{v}_m| = 1, \quad m = 1,\ldots,M,
\end{aligned}
\end{equation}
where
\begin{equation}
\begin{aligned}
\sqrt{1 + A_k |\mathbf{v}^\dagger \mathbf{u}_k|^2} 
&\geq g^{(t)} \\
&= \sqrt{1 + A_k |\mathbf{v}_t^\dagger  \mathbf{u}_k|^2} 
+ \frac{A_k}{\sqrt{1 + A_k |\mathbf{v}_t^\dagger \mathbf{u}_k|^2}} \\
&\quad \times \mathfrak{Re} \Big\{ (\mathbf{u}_k^\dagger  \mathbf{v}_t) (\mathbf{u}_k^\dagger  (\mathbf{v} - \mathbf{v}_t)) \Big\},
\end{aligned}
\end{equation}
and, $\mathbf{v}_t$ is the given point at the $t$-th iteration of the successive convex approximation \cite{Mu2021}, whose objective function is a convex function of $\mathbf{v}$. However, problem (\ref{eq:re_phase_opt}) remains non-convex due to the constant modulus constraint. To address this issue, we adopt a Riemannian manifold optimization framework. By characterizing the geometry of the associated manifold, problem (\ref{eq:re_phase_opt}) can be reformulated as
\begin{equation}
\begin{aligned}
\min_{\mathbf{v} \in \mathcal{M}_{\mathbf{v}}} \quad & 
\mathcal{G}_{\mathcal{M}}(\mathbf{v}) = \sum_{k=1}^{K} \Big[ |\bar{\gamma}_k|^2 B_k \mathbf{v}^\dagger \mathbf{D}_k \mathbf{v} - 2 g^{(t)} \cdot \mathfrak{Re}\{\bar{\gamma}_k^{\ast}\} \Big] \\
\text{s.t.} \quad & |\mathbf{v}_m| = 1, \quad m=1,\ldots,M,
\end{aligned}
\end{equation}
where $\mathcal{M}$ signifies the product manifold of dimension $M$.

Since the modulus of each element in $\mathbf{v}$ is $1$, therefore, the search space for $\mathbf{v}$ is given by the Cartesian product of $M$ unit circles in the complex plane, formulated as 
$\mathcal{M}_{\mathbf{v}} = \{\mathbf{v} \in \mathbb{C}^M : |\mathbf{v}_1| = |\mathbf{v}_2| = \cdots = |\mathbf{v}_M| = 1\}$.

The tangent space $T_{\mathbf{v}}\mathcal{M}_{\mathbf{v}}$ at point $\mathbf{v} \in \mathcal{M}_{\mathbf{v}}$ is formulated as
$T_{\mathbf{v}}\mathcal{M}_{\mathbf{v}} = \{\chi_{\mathbf{v}} \in \mathbb{C}^M : \mathfrak{Re} \{\chi_{\mathbf{v}} \odot \mathbf{v}^{\ast} \} = \mathbf{0}\}$,
where vector $\chi_{\mathbf{v}}$ denotes the tangent direction at the point $\mathbf{v}$, while $\mathbf{0}$ represents the zero vector with $M$ dimensions.  

The retraction on the product manifold $\mathcal{M}_{\mathbf{v}}$ is defined as
\begin{equation}
\mathrm{Rtrc}(\chi_{\mathbf{v}}) = \frac{\mathbf{v} + \chi_{\mathbf{v}}}{\|\mathbf{v} + \chi_{\mathbf{v}}\|_2},
\end{equation}
where $\mathrm{Rtrc}(\chi_{\mathbf{v}})$ maps the tangent vector $\chi_{\mathbf{v}}$ back onto $\mathcal{M}_{\mathbf{v}}$.

The transport operation of a tangent vector $\chi_{\mathbf{v}}$ from point $\mathbf{v}^k \in \mathcal{M}_{\mathbf{v}}$ to $\mathbf{v}^{k+1} \in \mathcal{M}_{\mathbf{v}}$ is expressed as
\begin{equation}
\mathrm{Trspt}_{\mathbf{v}^k \to \mathbf{v}^{k+1}}(\chi_{\mathbf{v}}) 
= \chi_{\mathbf{v}} - \mathfrak{Re}\!\left\{ \chi_{\mathbf{v}}^{\ast} \odot \mathbf{v}^{k+1} \right\} \odot \mathbf{v}^{k+1}.
\end{equation}

Finally, the Riemannian gradient of $\mathcal{G}_{\mathcal{M}_{\mathbf{v}}}(\mathbf{v})$ is given by
\begin{equation}
\mathrm{grad}\,\mathcal{G}_{\mathcal{M}}(\mathbf{v}) 
= G - \mathfrak{Re}\!\left\{ G \odot \mathbf{v}^{\ast}  \right\} \odot \mathbf{v},
\end{equation}
\hspace{0.3cm}where
\begin{equation} \label{eq:G_d}
\begin{aligned}
G &= 2\left(\sum_{k=1}^{K} |\bar{\gamma}_k|^2 B_k \mathbf{D}_k \mathbf{v} 
- \sum_{k=1}^{K} \mathfrak{Re}\{\bar{\gamma}_k^{\ast}\}\mathbf{d}_k \right),  \\
\mathbf{d}_k &= \frac{A_k}{\sqrt{1 + A_k |\mathbf{v}_t^\dagger \mathbf{u}_k|^2}} 
(\mathbf{u}_k^\dagger \mathbf{v}_t)\mathbf{u}_k.
\end{aligned}
\end{equation}

\begin{figure*}[t]
\centering
\includegraphics[width=1\textwidth]{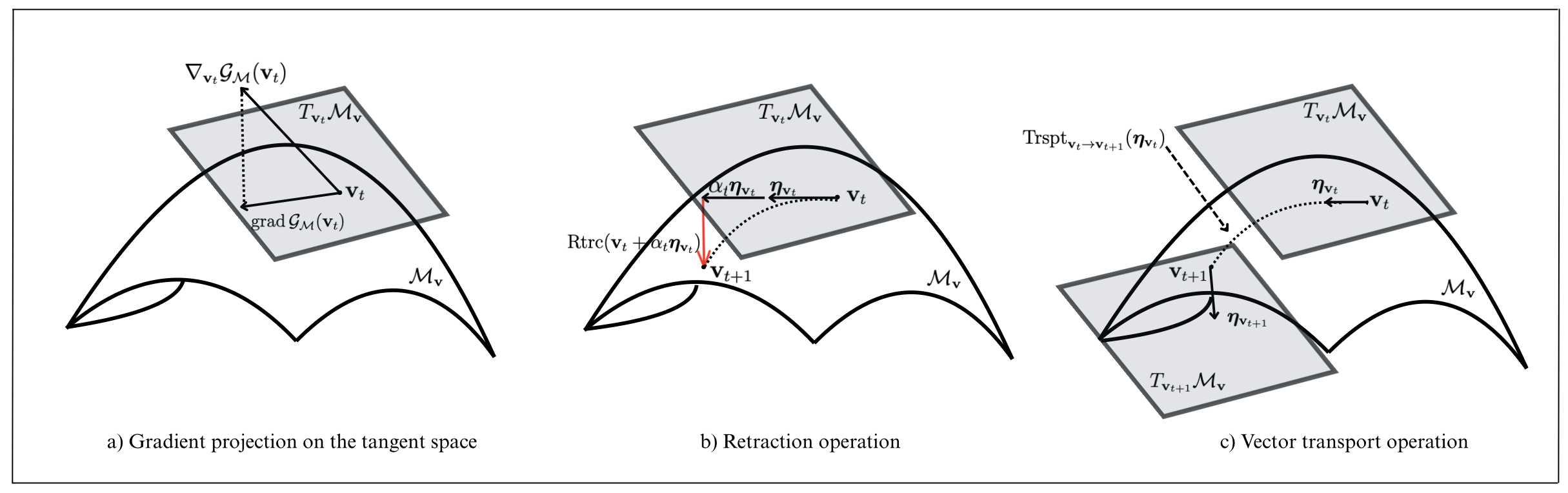}
\caption{Geometric illustration of the RCG process, (a) shows gradient projection, (b) shows retraction operation, and (c) shows vector transport operation on the manifold.}
\label{fig:RCG}
\vspace{-2mm}
\end{figure*}

The overall geometric process of the proposed RCG–based RIS phase–shift optimization is illustrated in Fig.~\ref{fig:RCG}. As shown in Fig.~\ref{fig:RCG}(a)–(c), the RCG procedure consists of three key operations: gradient projection onto the tangent space, retraction onto the manifold, and vector transport between successive tangent spaces, corresponding to the iterative updates summarized in Algorithm 2. 

\begin{algorithm}[ht]
\small
\caption{Riemannian Conjugate Gradient Optimization for RIS Phase Shifts}
\small
\begin{algorithmic}[1]
\State \textbf{Input:} $\mathbf{u}_k$, $\mathbf{D}_k$, $A_k$, $B_k$, initial RIS phase shifts: $\mathbf{v}_0 \in \mathbb{C}^M$, maximum outer iterations: $T_{\max}$, convergence tolerance: $\epsilon$.
\State \textbf{Output:} Optimized RIS phase shifts: $\mathbf{v}^{\text{opt}} $.
\State Initialize: $\mathbf{v} \leftarrow \mathbf{v}_0$, $t \leftarrow 0$;
\Repeat
    \For{$k = 1$ to $K$}
        \State Compute $\gamma_k^{\mathrm{opt}}$ and $\mathbf{d}_k$ according to (\ref{eq:gamma_opt}) and (\ref{eq:G_d});
    \EndFor
    \State Compute Euclidean Gradient according to (\ref{eq:G_d});
    \State Calculate the Riemannian gradient:
    \State $\boldsymbol{\xi} \leftarrow G - \mathfrak{Re}\{ G \odot \mathbf{v}^{\ast}  \} \odot \mathbf{v}$
    \State Initialize the search direction and the counter:
    \State $\boldsymbol{\eta} \leftarrow -\boldsymbol{\xi}, \quad j \leftarrow 0$
    \While{$\|\boldsymbol{\xi}\| > \epsilon$ and $j < J_{\max}$}
        \State Determine step size $\alpha$ via Armijo line search;
        \State Update RIS phase shifts:
        \State $\mathbf{v}_{\mathrm{new}} = \mathrm{Rtrc}(\mathbf{v} + \alpha \boldsymbol{\eta})$
        \State Update Euclidean Gradient $G_{\mathrm{new}}$ according to (\ref{eq:G_d});
        \State Update Riemannian gradient $\boldsymbol{\xi}_{\mathrm{new}}$ according to (\ref{eq:G_d});
        \State Calculate the PRP conjugate coefficient:
        \State $\bar{\beta} = \dfrac{\langle \boldsymbol{\xi}_{\mathrm{new}}, \boldsymbol{\xi}_{\mathrm{new}} - \boldsymbol{\xi} \rangle}{\langle \boldsymbol{\xi}, \boldsymbol{\xi} \rangle}$
        \State Calculate the new conjugate direction:
        \State $\boldsymbol{\eta}_{\mathrm{new}} = -\boldsymbol{\xi}_{\mathrm{new}} + \bar{\beta}\left[\boldsymbol{\eta} - \mathfrak{Re}\{\boldsymbol{\eta} \odot \mathbf{v}_{\mathrm{new}}^{\ast}\} \odot \mathbf{v}_{\mathrm{new}} \right]$
        \State Variable update and iteration counting:
        \State $\mathbf{v} \leftarrow \mathbf{v}_{\mathrm{new}}, \;
        \boldsymbol{\xi} \leftarrow \boldsymbol{\xi}_{\mathrm{new}}, \;
        \boldsymbol{\eta} \leftarrow \boldsymbol{\eta}_{\mathrm{new}}, \;
        j \leftarrow j + 1$
    \EndWhile
    \State Update the counter: $t \leftarrow t+1$;
\Until{$|R_s(\mathbf{v}) - R_s(\mathbf{v}_{\mathrm{prev}})| < \epsilon$ or $t > T_{\max}$}
\State \Return $\mathbf{v}$.
\end{algorithmic}
\end{algorithm}

\subsection{Computational Complexity Analysis}
We now analyze the computational complexity of Algorithm~1. Let $M$ denote the number of RIS elements, $K$ the number of users, $L$ the maximum number of trials in the Armijo line search per inner iteration, $J$ the maximum number of inner iterations per outer iteration, and $T$ the maximum number of outer iterations. The complexity of each step is characterized as follows.

\begin{itemize}
    \item \textbf{Computation of $\gamma_k^{\mathrm{opt}}$ and $\mathbf{d}_k$:}  
    As shown in (\ref{eq:gamma_opt}) and (\ref{eq:G_d}), the evaluation of $|\mathbf{v}^\dagger \mathbf{u}_k|^2$ requires $\mathcal{O}(M)$ operations per user, while the quadratic form $\mathbf{v}^\dagger \mathbf{D}_k \mathbf{v}$ incurs $\mathcal{O}(M^2)$. Across all $K$ users, this step yields $\mathcal{O}(KM^2)$.

    \item \textbf{Computation of the Riemannian gradient:}  
    Each matrix–vector multiplication $\mathbf{D}_k \mathbf{v}$ requires $\mathcal{O}(M^2)$, repeated for $K$ users, giving $\mathcal{O}(KM^2)$. The summation terms add $\mathcal{O}(KM)$, but the overall complexity remains $\mathcal{O}(KM^2)$.

    \item \textbf{Armijo line search:}  
    Each trial involves recomputation of the objective function, dominated by $\mathbf{v}^\dagger \mathbf{D}_k \mathbf{v}$, leading to $\mathcal{O}(KM^2)$ per trial. With at most $L$ trials, this step incurs $\mathcal{O}(LKM^2)$ per inner iteration.

    \item \textbf{Retraction and conjugate direction update:}  
    Both retraction $\mathbf{v}_{\mathrm{new}} = \mathrm{Rtrc}(\mathbf{v} + \alpha \boldsymbol{\eta})$ and update of the conjugate direction require $\mathcal{O}(M)$ operations.

    \item \textbf{Convergence check:}  
    The secrecy rate calculation requires evaluating $R_k$ for all $K$ users, each costing $\mathcal{O}(M^2)$, leading to $\mathcal{O}(KM^2)$ per outer iteration.
\end{itemize}

The per–inner–iteration complexity is summarized in Table~\ref{tab:complexity}.  

\begin{table}[h]
\centering
\caption{Total complexity (per Inner Iteration)}
\label{tab:complexity}
\begin{tabular}{lc}
\hline
\textbf{Step} & \textbf{Complexity} \\
\hline
Computation of $\gamma_k^{\mathrm{opt}}$, $\mathbf{d}_k$ & $\mathcal{O}(KM^2)$ \\
Computation of Riemannian gradient & $\mathcal{O}(KM^2)$ \\
Armijo line search & $\mathcal{O}(LKM^2)$ \\
Update of $\mathbf{v}$ and $\boldsymbol{\eta}$ & $\mathcal{O}(M)$ \\
\hline
\textbf{Total per inner iteration} & $\mathcal{O}(LKM^2)$ \\
\hline
\end{tabular}
\end{table}

Considering $T$ outer iterations and $J$ inner iterations, the complexity per inner iteration is $\mathcal{O}(LKM^2)$. 
Since the inner loop executes at most $J$ times within each outer iteration, and the outer loop runs for a maximum of $T$ iterations, the overall computational burden can be expressed as
\begin{equation}
\mathcal{O}\!\left(T J L K M^2 \right) + \mathcal{O}\!\left(T K M^2 \right)
= \mathcal{O}\!\left(T J L K M^2 \right),
\label{eq:complexity}
\end{equation}
where the second term, arising from the convergence check, is negligible compared to the first.

It is worth noting that the proposed algorithm is well-suited for scenarios involving moderate-scale RIS deployments (e.g., $M$ up to a few hundred) and a moderate number of users $K$. However, when the number of RIS elements becomes very large (e.g., $M \geq 1000$), the $\mathcal{O}(M^2)$ complexity might become prohibitive, and approximate methods or simplified models might be necessary to reduce the computational burden.

\section{Simulation Results}
In this section, we evaluate the performance of the proposed RIS-assisted CFmMIMO system through extensive Monte Carlo simulations. A single AP equipped with $N_t=16$ transmit antennas serves two legitimate users, and a single Eve. Each Bob is assigned an exclusive subset of $n_t=8$ transmit antennas and employs a single receive antenna, while Eve is equipped with $N_E=3$ receive antennas. Due to blockage, no direct AP–UE links are considered. The AP is assumed to have perfect knowledge of the cascaded AP–RIS–UE channels for analysis. Unless otherwise specified, the RIS consists of $M = 16$ reflecting elements with unit-modulus phase shifts, the total transmit power is normalized to $\mathcal{P}_{\mathrm{tot}} = 1$, and each performance metric is averaged over at least $10^{4}$ independent Monte Carlo realizations. The convergence tolerance is set to $\epsilon = 10^{-2}$, and quadrature amplitude modulation (QAM) is employed for data transmission. The RIS-Bob and RIS-Eve channels follow a distance-dependent large-scale fading model of the form
\begin{equation}\label{eq:channel}
    \mathbf{H}_{\ell} = \sqrt{L_0 d_{\ell}^{-\alpha}}\,\widetilde{\mathbf{H}}_{\ell},
\end{equation}
where $L_0 = (\lambda_c / 4\pi)^2$ denotes the reference path-loss constant with $\lambda_c$ being the wavelength of the center frequency of the information carrier. $\alpha$ is the path-loss exponent corresponding to outdoor propagation, and $\widetilde{\mathbf{H}}_{\ell}$ is a matrix of i.i.d. $\mathcal{CN}(0,1)$ small-scale fading coefficients.

We evaluate four transmission strategies to benchmark the proposed RIS-assisted CFmMIMO system. The baseline scheme (BS) represents a fully non-optimized configuration in which both the beamforming matrix $\mathbf{W}$ and the RIS phase-shift matrix $\boldsymbol{\Theta}$ are generated randomly, and equal power allocation is applied. The RCG RIS optimization scheme isolates the impact of RIS configuration by optimizing $\boldsymbol{\Theta}$ via the Riemannian conjugate gradient algorithm, while keeping $\mathbf{W}$ random and maintaining equal power allocation. The joint beamforming and RCG RIS optimization scheme improves upon this by jointly optimizing the beamforming matrix and the RIS phase-shift matrix, while maintaining equal power allocation across users. Finally, the alternating optimization (AO) scheme performs a unified optimization of $\mathbf{W}$, $\boldsymbol{\Theta}$, and the power allocation vector, thereby fully exploiting spatial and power-domain degrees of freedom to maximize the achievable secrecy rate.

\begin{figure}[t]
\centerline{\includegraphics[width=80mm]{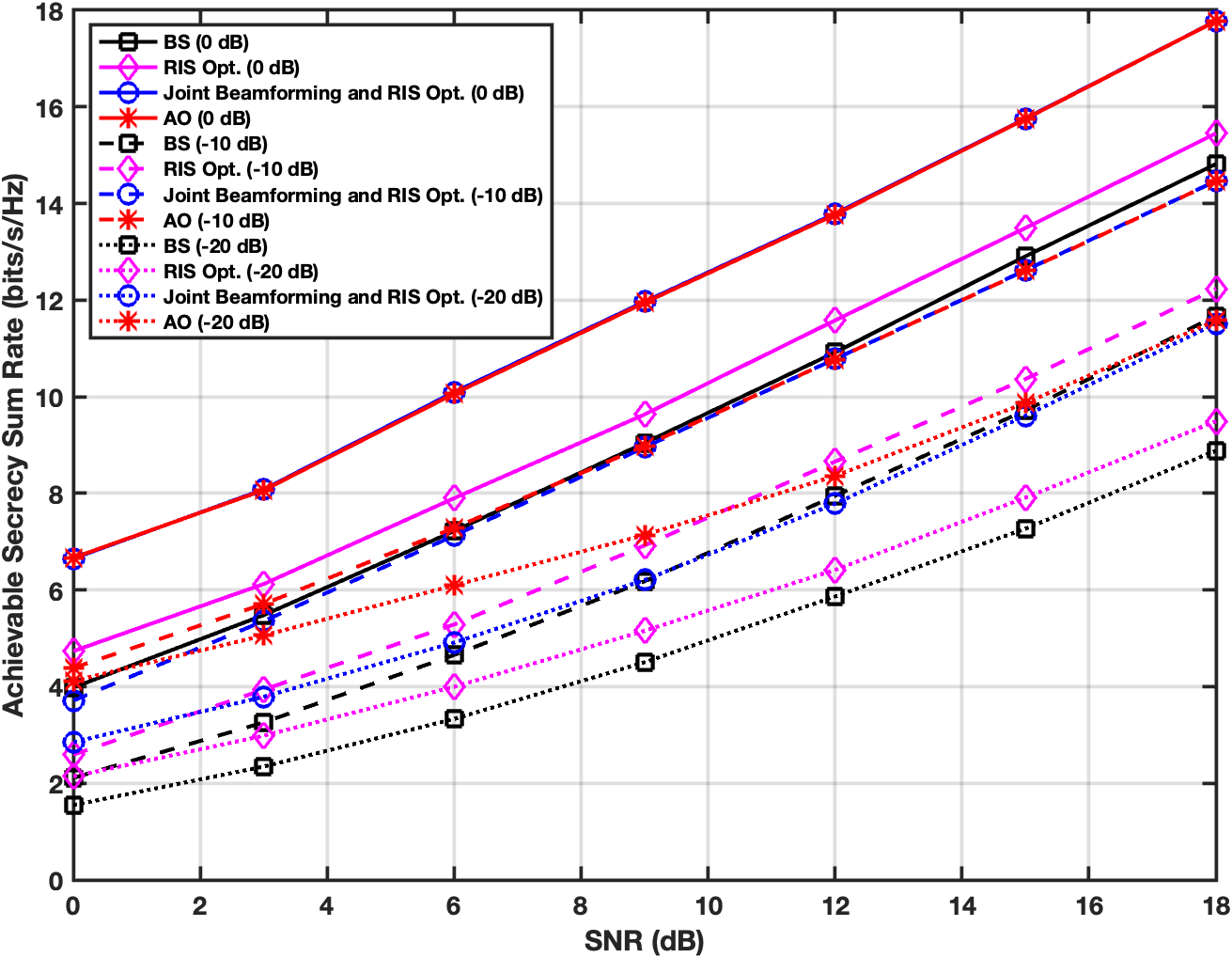}}
\caption{Achievable secrecy sum rate versus SNR for the proposed and baseline schemes.}
\label{fig2}
\end{figure}

Fig.~\ref{fig2} presents the achievable secrecy sum rate versus SNR for the alternating optimization scheme, along with three comparison schemes: the baseline scheme, RIS optimization, and joint beamforming and RIS optimization, evaluated under three different large-scale gain conditions for Bob2. These scenarios correspond to large-scale fading differences of 0 dB, -10 dB, and -20 dB relative to Bob1. Across all SNR values and all three channel gain configurations, the alternating optimization strategy consistently achieves the highest secrecy sum rate. When Bob2 experiences no gain disadvantage (0 dB), both users have identical large-scale fading, so the power allocation step remains unchanged and splits power symmetrically. In this balanced setting, the comparison schemes do not suffer from user imbalance, and the alternating optimization method exhibits a more pronounced advantage because the optimized beamforming and RIS phases provide additional gains that the other schemes cannot achieve. As the disparity increases to -10 dB, the comparison schemes begin to exhibit noticeable secrecy degradation due to their inability to simultaneously manage interference and suppress leakage toward Eve. At the same time, the alternating optimization design remains robust thanks to its coordinated optimization of beamforming, RIS phases, and power distribution. Under the most challenging case (-20 dB), the comparison schemes suffer a pronounced drop in secrecy performance, particularly at high SNR. In contrast, the alternating optimization scheme continues to provide secure communication by effectively prioritizing the stronger user and shaping the cascaded channels through the optimized RIS configuration. It is also observed that optimizing the AP beamformer yields the dominant performance gain, as the beamforming design constitutes a convex subproblem that leads to the global optimum, aligning the effective channels and suppressing inter-user interference. The RIS phase-shift design, being non-convex, converges only to a local optimum and thus contributes a smaller but consistent improvement.
\begin{figure}[t]
\centerline{\includegraphics[width=80mm]{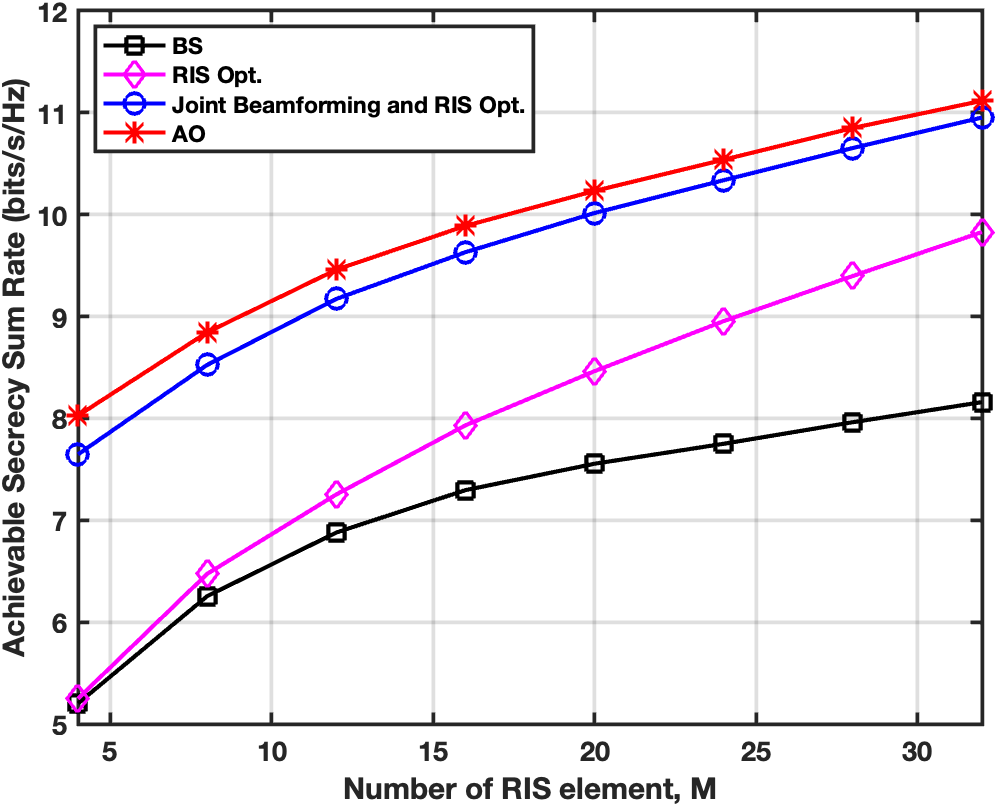}}
\caption{Average secrecy sum rate versus number of RIS elements.}
\label{fig3}
\end{figure}

Fig. \ref{fig3} illustrates the achievable secrecy sum rate versus the number of RIS elements $M$ for the alternating optimization scheme and the three comparison schemes. As expected, increasing $M$ consistently enhances the secrecy performance across all schemes since a larger RIS surface provides stronger array gains and greater flexibility in shaping the cascaded channels. The alternating optimization scheme achieves the highest secrecy sum rate for all values of $M$, demonstrating the benefit of simultaneously optimizing the transmit beamforming, RIS phase shifts, and power allocation. Among the comparison schemes, the gap between the baseline scheme and the RIS optimization scheme highlights the improvement obtained through optimizing only the RIS phase-shift matrix. At the same time, the additional gain provided by the joint beamforming and RIS optimization scheme reflects the contribution of optimizing the active beamformer. The performance hierarchy becomes increasingly visible as $M$ grows. When $M$ is small, all schemes are constrained by limited reflection capability, resulting in modest rate differences. However, for larger RIS sizes, the alternating optimization scheme yields substantial gains because the optimized RIS-AP coordination more effectively enhances the legitimate links while suppressing information leakage toward Eve.
\begin{figure}[t]
\centerline{\includegraphics[width=78mm]{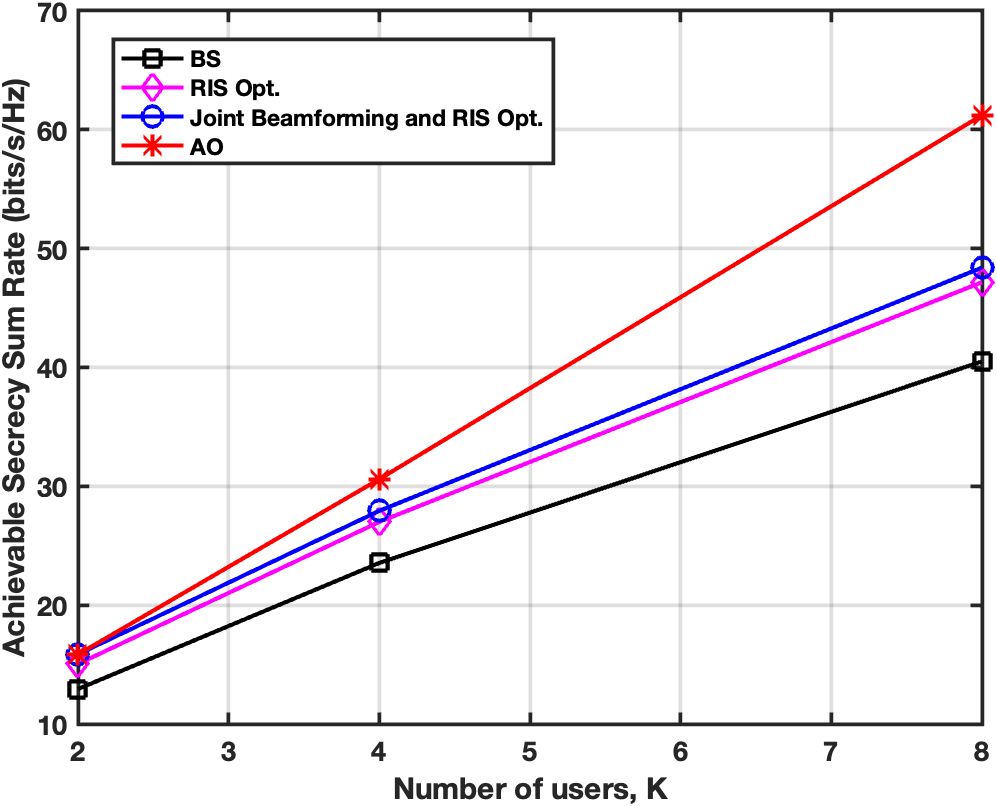}}
\caption{Average secrecy sum rate versus number of legitimate users.}
\label{fig4}
\end{figure}

Fig. \ref{fig4} illustrates the achievable secrecy sum rate as the number of users, $K$, increases, while keeping the number of transmit antennas per user fixed at $n_t = 8$ and the RIS size fixed at $M_{\mathrm{RIS}} = 16$. Thus, the total number of AP antennas scales linearly with $K$ according to $N_t = K n_t$. The alternating optimization scheme consistently achieves the highest secrecy sum rate for all user loads, demonstrating its robustness in multi-user settings. Although all schemes exhibit increasing secrecy sum rate with larger $K$ as more users introduce additional independent data streams, the growth rates differ noticeably. The baseline scheme yields the lowest performance since neither the beamforming nor the RIS phase shifts are optimized, resulting in increased interference as $K$ increases. The RIS optimization and the joint beamforming and RIS optimization schemes achieve higher secrecy rates due to improved control of the propagation environment, but both lack coordinated power adaptation and therefore experience performance degradation more rapidly in denser networks.
\begin{figure}[t]
\centerline{\includegraphics[width=80mm]{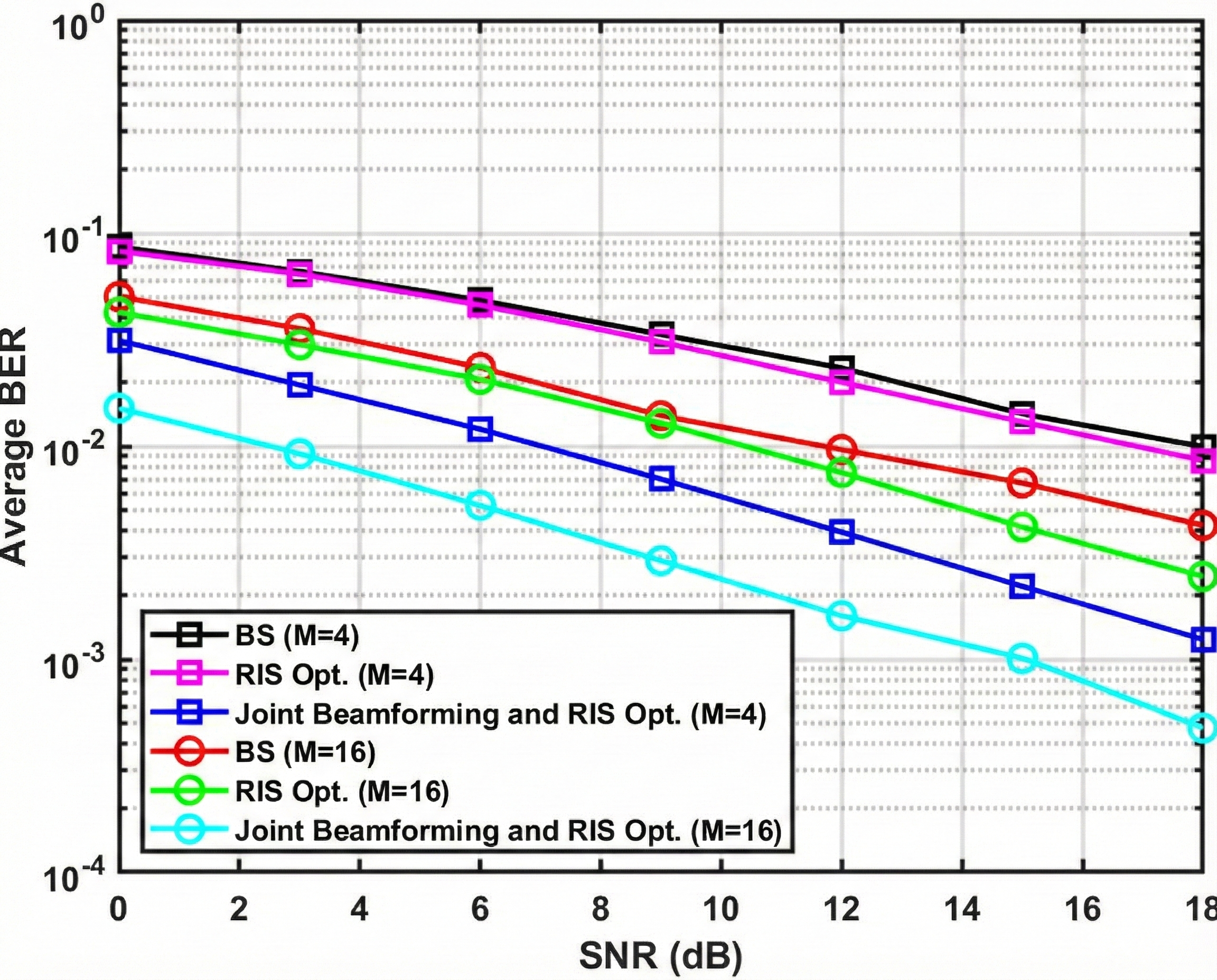}}
\caption{Average BER of Bob 2 versus SNR for RIS sizes $M=4$ and $M=16$ under a $0$ dB large-scale channel gain.}
\label{fig5}
\end{figure}

Fig. \ref{fig5} presents the average BER performance of the three transmission strategies considered, namely the baseline scheme, RIS optimization, and joint beamforming and RIS optimization, for the number of RIS elements $M=4$ and $M=16$, evaluated for Bob 2 with a large-scale channel gain of $0$ dB. Since all users experience identical distance-dependent path loss, power optimization is unnecessary, and analyzing Bob 2 is sufficient. For both RIS configurations, the baseline scheme yields the highest BER because neither the beamforming matrix nor the RIS phase shifts are optimized. The RIS optimization scheme provides a moderate BER reduction relative to the baseline, reflecting the benefit of configuring only the RIS phase-shift matrix while keeping the beamforming matrix unchanged. The joint beamforming and RIS optimization scheme achieves the most substantial performance improvement across all SNR values, as the coherent design of both active and passive beamforming components significantly enhances the effective channel of Bob 2 even without power allocation. As expected, increasing the number of RIS elements from $M=4$ to $M=16$ uniformly shifts the BER curves downward due to the improved array gain and enhanced passive beamforming capability.

\begin{figure}[b]
\centerline{\includegraphics[width=80mm]{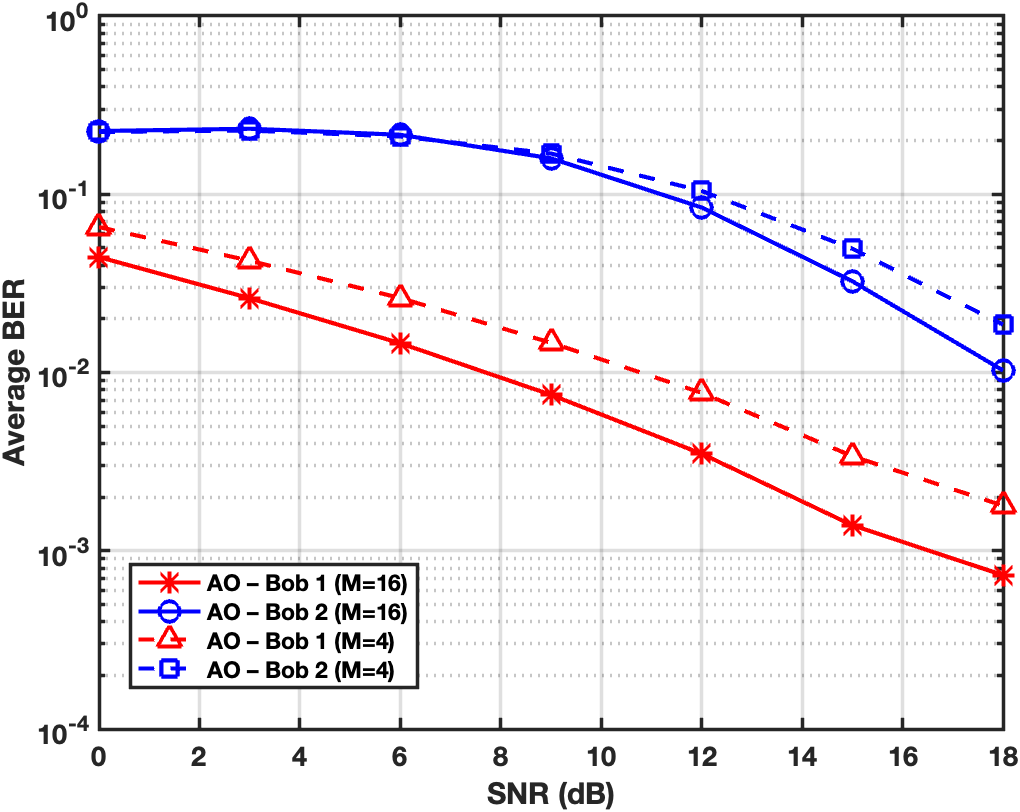}}
\caption{Average BER performance versus SNR of the alternating optimization scheme for Bob 1 and Bob 2 with large-scale fading of 0 dB and -20 dB, respectively, using $M=4$ and $M=16$ RIS elements.}
\label{fig6}
\end{figure}

Fig. \ref{fig6} illustrates the BER performance of the AO scheme for Bob 1 with 0 dB large-scale fading and Bob 2 with -20 dB large-scale fading, using $M = 4$ and $M = 16$ RIS elements. As expected, Bob 1 achieves much lower BER across all SNR values due to its stronger large-scale channel, while Bob 2 exhibits higher BER because of its severe path loss disadvantage. Increasing the number of RIS elements improves the BER behavior for both users, particularly at moderate and high SNR values. Although Bob 2 remains performance-limited, the alternating optimization procedure still ensures that both the beamformer and RIS phases adapt jointly to the channel conditions, enabling each user to operate as reliably as possible under its respective link quality.

\section{Conclusion}
In this paper, we have presented a comprehensive optimization framework for RIS–assisted CFmMIMO networks, focusing on enhancing physical layer security and mitigating multi-user interference. A CSI-based precoder was developed at the access point to suppress interference among users, enabling interference-free reception for legitimate users while inherently generating jamming toward the eavesdropper. The secrecy sum rate maximization problem was addressed through an alternating optimization framework that jointly refines the beamforming vectors, user power allocation, and RIS phase-shift matrix. The RIS phase-shift optimization was performed via a RCG algorithm on the complex unit-modulus manifold, achieving efficient convergence with an overall computational complexity of $\mathcal{O}(T J L K M^2)$. Simulation results verified that the proposed framework substantially improves secrecy performance and interference suppression compared with baseline schemes, confirming its effectiveness for secure and scalable CFmMIMO systems in dense wireless environments.
\section{Appendix A}
To simplify the notation in the proof, define the following: $\mathbf{\xi} \stackrel{\Delta}{=} \bar{\beta} ( \mathbf{a}-\mathbf{b})$, where $ \mathbf{a}= \mathbf{G} \mathbf{W}_{\mathrm{opt}} \mathbf{S} \bf\varpi_{opt}$ and $\mathbf{b}= \mathbf{G}_{k} \mathbf{w}_k^{\mathrm{opt}} s_{k}.$ Then, the jamming vector  in (\ref{eq:jamming}) can be written as $\mathbf{J}^{(k)}_E=\bar{\beta} \mathbf{H}_E \mathbf{\Theta}\mathbf{\xi}$ and consequently the jamming variance can be expressed as
 \begin{eqnarray}\label{app1}
 	\sigma^{2}_{\mathbf{J}^{(k)}_E }&=& \mathbb{E}_{\mathbf{H}_E, \mathbf{G}}\{\mathbf{J}^{(k)\dagger}_E
 	 \mathbf{J}^{(k)}_E\} \nonumber\\
 	&=& \bar{\beta}^2 \mathbb{E}_{\mathbf{H}_E, \mathbf{G}}\left\{\mathbf{\xi}^{\dagger} \mathbf{\Theta}^{\dagger}\mathbf{H}^{\dagger}_{E} \mathbf{H}_E \mathbf{\Theta}\mathbf{\xi}\right\},
 \end{eqnarray}
where $\mathbf{H}_E \in \mathcal{C}^{M \times N_E}$ is the channel matrix between the source and the eavesdropper with independent and identically distributed $(\text{i.i.d.})$ entries according to $\mathcal{CN}(0,1)$. Therefore, it can be easily seen that $\mathbb{E}_{\mathbf{H}_E}\left\{\mathbf{H}^{\dagger}_{E} \mathbf{H}_E \right\}=N_E \mathbf{I}_M$, where $\mathbf{I}_M$ denotes a $M \times M$ diagonal unit matrix. Since also $\mathbf{\Theta}^{\dagger}\mathbf{\Theta}=\mathbf{I}_M$, it follows from (\ref{app1}) that 
\begin{eqnarray}\label{app2}
	\sigma^{2}_{\mathbf{J}^{(k)}_E }&=&\bar{\beta}^2 N_E \mathbb{E}_{\mathbf{H}_E, \mathbf{G}}\left\{\mathbf{\xi}^{\dagger} \mathbf{\xi}\right\}\\
	&=& \bar{\beta}^2 N_E \mathbb{E}_{\mathbf{H}_E, \mathbf{G}}\left( \parallel \mathbf{a}\parallel^{2}+\parallel \mathbf{b}\parallel^{2}-2 R\{\mathbf{b}^{\dagger}\mathbf{a } \}\right) \nonumber.
\end{eqnarray}
Since the channel matrix $\mathbf{G}\in \mathcal{C}^{M \times N_t}$ between the AP-RIS has also $(\text{i.i.d.})$ entries according to $\mathcal{CN}(0,1)$ it follows that $\mathbb{E}_{\mathbf{G}}\left\{\mathbf{G}^{\dagger} \mathbf{G} \right\}=\!\!M  \mathbf{I}_{N_t}$. In addition, taking into facts that $\parallel\mathbf{W}_{\mathrm{opt}}\parallel^{2} =\mathcal{P}\mathbf{I}_K$, and $\parallel\mathbf{ s}\parallel^2=1$ as well as  from (\ref{eq:beta}) it follows that $\bf \varpi^\dagger_{opt}\bf \varpi_{opt}=K/\bar{\beta}^2$,  each term in (\ref{app2}) can be computed as
\begin{eqnarray}
\mathbb{E}_{ \mathbf{G}}\left( \parallel \mathbf{a}\parallel^{2}\right)&=& \frac{M\mathcal{P}K}{\bar{\beta}^2}, \label{eq:31}\\
\mathbb{E}_{ \mathbf{G}}\left( \parallel \mathbf{b}\parallel^{2}\right)&=& M\mathcal{ P}\label{eq:32},\\
\mathbb{E}_{ \mathbf{G}} \left(\mathbf{b}^{\dagger} \mathbf{a} \right)&=& M \varpi_{opt} (k). \label{eq:33}
\end{eqnarray}
Finally substituting (\ref{eq:31}), (\ref{eq:32}) and (\ref{eq:33}) into (\ref{app2}), the final result follows as given in the \textit{Lemma 1}.

\end{document}